\documentclass[11pt]{article}
\usepackage[utf8]{inputenc}
\usepackage{amsmath,setspace,geometry}
\usepackage{amsthm}
\usepackage{amsfonts}
\usepackage[shortlabels]{enumitem}
\usepackage{rotating}
\usepackage{pdflscape}
\usepackage{graphicx}
\usepackage{bbm}
\usepackage{comment}
\usepackage[dvipsnames]{xcolor}
\usepackage{hyperref}
\hypersetup{colorlinks=true, linkcolor= BrickRed, citecolor = BrickRed, filecolor = BrickRed, urlcolor = BrickRed, hypertexnames = true}
\usepackage[]{natbib} 
\bibpunct[:]{(}{)}{,}{a}{}{,}
\usepackage[english]{babel}
\usepackage{float}
\usepackage{caption}
\usepackage{subcaption}
\usepackage{booktabs}
\usepackage{pdfpages}
\usepackage{threeparttable}
\usepackage{lscape}
\usepackage{bm}
\title{Nonparametric Estimation of Matching Efficiency and Elasticity in a Spot Gig Work Platform: 2019–2023}
\author{Hayato Kanayama\thanks{\href{mailto:}{ha13yato@toki.waseda.jp} Graduate School of Economics, Waseda University.}, Suguru Otani\thanks{\href{mailto:}{suguru.otani@e.u-tokyo.ac.jp}, Market Design Center, Department of Economics, The University of Tokyo\\
 We thank Fuhito Kojima, Kosuke Uetake, Akira Matsushita, Kazuhiro Teramoto, Masayuki Sawada, Shoya Ishimaru, Ryo Kambayashi, Higashi Yudai, Sho Miyaji, and Masaru Sasaki for their valuable advice. We acknowledge Kazuki Takaishi and Takahide Kimura for sharing the data and their technical and institutional knowledge in the Timee platform. We also thank seminar and conference participants at Hitotsubashi University, Waseda University, Summer Workshop on Economic Theory (SWET), and 27th Labor Economics Conference at University of Tokyo. This work was supported by JST ERATO Grant Number JPMJER2301, and 25K16620, Japan. } }
 
\date{
First version: December 26, 2024\\
Current version: \today
}

\begin{document}

\maketitle

\begin{abstract}
    This paper provides new evidence on spot gig work platforms for individuals seeking flexible, short-term jobs with minimal educational or experience requirements in Japan. Using proprietary data from Timee, a private matching platform, the study analyzes trends in active users, vacancies, hires, and labor market tightness, compared to part-time data from Hello Work, a public employment service. Applying a nonparametric approach, it finds that the private platform exhibits substantially higher matching efficiency, especially after 2022. Elasticities also differ across platforms: for Hello Work, the user elasticity fluctuates around 0.3--0.5, while the vacancy elasticity ranges roughly from 0.4 to slightly above 1.0; for the private platform, the user elasticity remains around 0.2--0.3, while the vacancy elasticity ranges from 0.7 to 1.1. At the prefecture level, the three prefectures exhibit broadly similar movements early in the sample, followed by divergence and partial re-convergence later on, while elasticities remain stable and similar across regions. These results reveal how digital platforms reshape job matching dynamics relative to traditional systems.
    \quad \\
    %100 words AER (now 93 words)
\end{abstract}

\section{Introduction}
The gig economy, characterized by temporary, contract, and freelance online jobs rather than permanent positions, has experienced rapid growth in recent years. An increasing proportion of freelance workers are now matched with customers through online platforms. By analyzing the number of open projects and tasks on a sample of such platforms, \cite{kassi2018online} find that the demand for online gig work increased by approximately 21\% from 2016 to 2018. According to \textit{Freelance Forward 2023} reported by Upwork, the share of professionals freelancing increased to nearly 64 million Americans, making up 38\% of the U.S. workforce in 2023.\footnote{\url{https://www.upwork.com/research/freelance-forward-2023-research-report}: Accessed on September 19, 2024.} Despite the significance of the gig economy, research directly analyzing the structure and efficiency of gig labor markets remains limited, especially in comparison to traditional labor markets facilitated by public job search platforms.

\begin{figure}[!ht]
  \begin{center}
  \subfloat[Hires, Labor force survey]{\includegraphics[width = 0.6\textwidth]
  {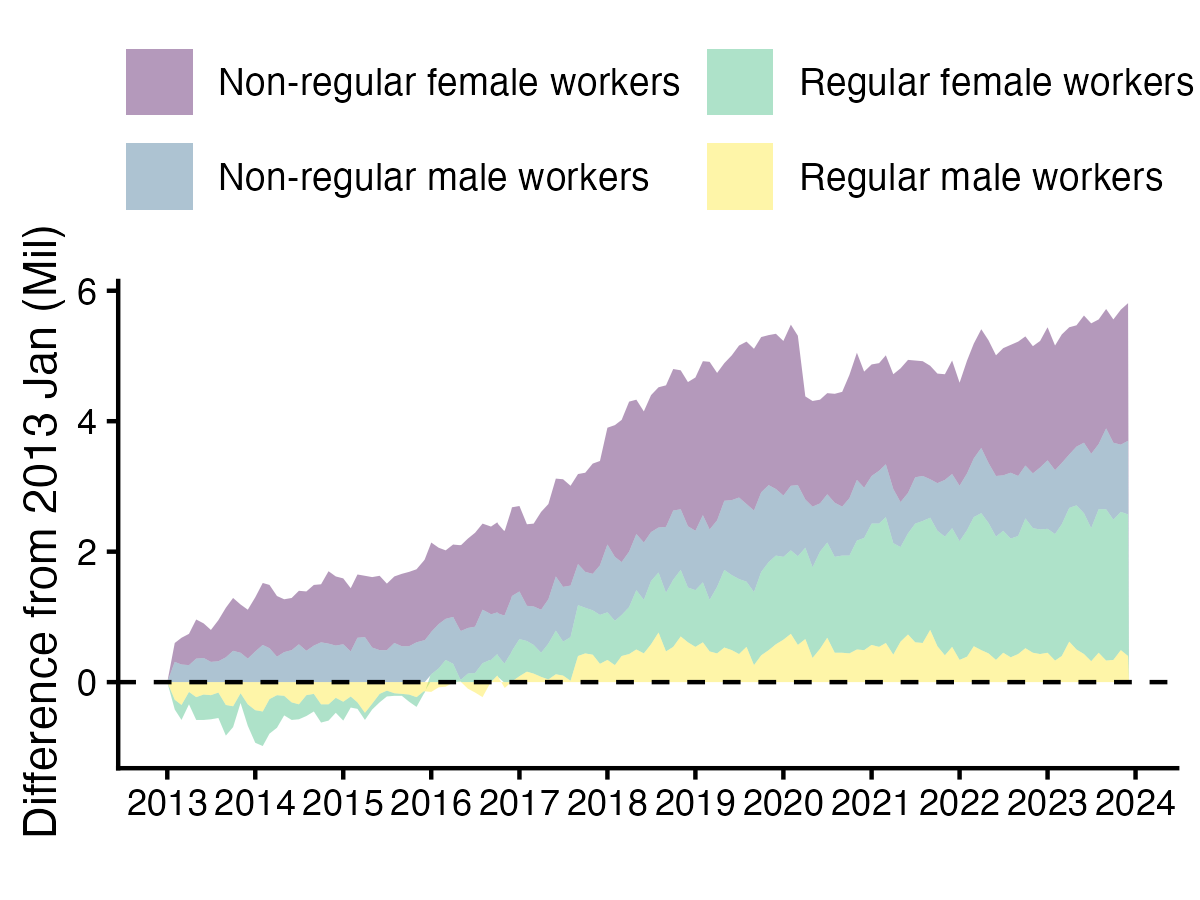}}\\
  \subfloat[Hires, Spot work platform]{\includegraphics[width = 0.6\textwidth]
  {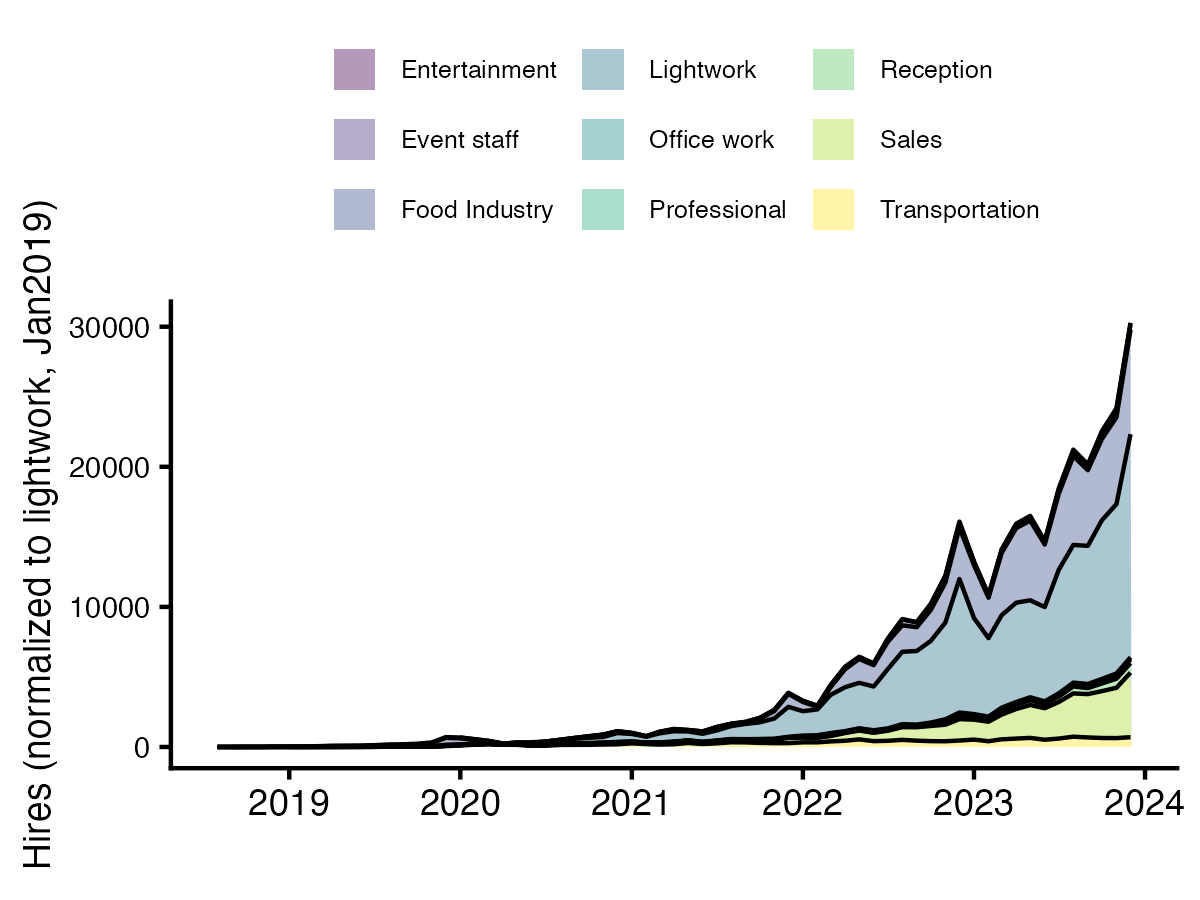}}
  \caption{Changes of employment. National vs Spot work platform}
  \label{fg:labor_force} 
  \end{center}
  \footnotesize
  Note: Panel (a) illustrates the time series of job changes in Japan, standardized to January 2013. The colored areas represent changes within specific worker groups. This graph is derived from data provided by the Economic Survey conducted by the Japanese Cabinet Office (details available at \url{https://www5.cao.go.jp/keizai3/2024/0802wp-keizai/setsumei-e2024.pdf}, accessed September 19, 2024). Panel (b) presents the time series of hires on the platform, standardized to January 2019 for confidentiality purposes. The vertical axis represents the relative number of hires compared to January 2019. The filled areas depict the number of matched spot workers across various job categories as defined by the respective companies. See the survey of \cite{miyamoto2025macroeconomic} for an overview of labor markets in Japan.
\end{figure}

This paper contributes to the literature by providing the first systematic analysis of a spot gig worker labor market, distinguishing it from traditional labor markets and other segments of the gig economy. We focus on labor market matching efficiency and elasticities within spot work platforms that cater to workers who want to work as an on-demand labor force or are unemployed, seeking flexible, short-term jobs requiring minimal education or experience. Using proprietary aggregate data from a private online spot work matching platform, Timee, in Japan, we document key trends and compare these to public-sector employment services. The significance of spot work markets has expanded considerably from 2013 to 2024, as shown in Panels (a) and (b) of Figure \ref{fg:labor_force}. First, we analyze trends in core aggregate variables, including the number of Timee users, vacancies, hires, and labor market tightness (i.e., the ratio of vacancies to Timee users). To benchmark our findings against the public-sector labor market, we incorporate data from the ``Report on Employment Service" (\textit{Shokugyo Antei Gyomu Tokei}), which provides insights into part-time employment trends facilitated through Hello Work, Japan's public employment service. This comparison sheds light on the structural differences not only between private and public job-matching mechanisms but also between gig and standard jobs in Japan.

First, we document the rapid expansion of the spot labor market on the private platform between December 2019 and December 2023. The framework follows \cite{cullen2021outsourcing} which studies a more housework-specific spot job platform, TaskRabbit. The number of registered Timee users, predominantly unemployed individuals actively seeking spot work, exhibits a gradual increase, while the number of vacancies surges—particularly after 2022—leading to a rise in labor market tightness. Hiring activity escalates significantly around 2020, reflecting a sharp rise in successful job matches. Additionally, the job finding rate surpasses 1.0, indicating that active users frequently secure multiple spot jobs per month. Meanwhile, the worker finding rate remains stable at around 0.8, showing that 80\% of spot jobs are successfully filled, reinforcing the platform's effectiveness in facilitating rapid job matching and featuring the matching frequency of the spot gig work platform.

Second, we estimate the matching function and recover matching efficiency and elasticities using the novel nonparametric approach developed by \cite{lange2020beyond}. The approach is applicable to a wide range of matching markets, including full-time labor and marriage markets \citep{otani2024nonparametric,otani2025onthejob, otani2025marriage}. Our results reveal clear distinctions between public and private job-matching mechanisms. For Hello Work, matching efficiency remains relatively stable, fluctuating within a narrow range around the baseline. In contrast, the private platform exhibits a substantial rise in matching efficiency, peaking around May 2023, coinciding with Japan's peak holiday season, Golden Week. Elasticities also differ: for Hello Work, the elasticity with respect to users fluctuates around 0.3 to 0.5, while the elasticity with respect to vacancies ranges roughly from 0.4 to slightly above 1.0. For the private platform, the elasticity with respect to users remains relatively low, around 0.2 to 0.3, whereas vacancy elasticity ranges from 0.7 to 1.1, indicating stronger responsiveness to job demand than to worker inflows.

\textcolor{black}{These comparisons should be interpreted based on differences in the measurement of variables and the composition of workers, vacancies, and hires across the two datasets, as well as the substantive difference between more standard job matching through Hello Work and spot gig work on an app-based platform. Estimating a matching function for both Timee and Hello Work is conceptually appropriate because both settings involve hires, job seekers, and vacancies, but differences in how these objects are measured and constituted affect how the estimated efficiency and elasticities should be interpreted. Accordingly, differences between the two platforms should be understood as reflecting not only differences in matching mechanisms, but also differences in the underlying data structure and in the type of jobs being matched. At the same time, this comparison provides broader insight into how gig work can be incorporated into search-and-matching models and how app-based spot labor markets differ from more traditional matching environments.}

\textcolor{black}{Lastly, we examine prefecture-level trends, focusing on Aichi, Osaka, and Tokyo---Japan's three largest metropolitan areas. This comparison is intended to assess whether the platform exhibits similar matching performance across major regions, thereby shedding light on geographic heterogeneity in spot-work matching on an app-based platform. We focus on these prefectures for two reasons. First, at the prefecture level, many smaller prefectures are too thin, especially early in the sample, for spot-work matching to form a meaningful market-level series. Second, \cite{kanayama2026justminimumwagehikes} documents that Timee first launched in major metropolitan prefectures such as Tokyo, Osaka, and Aichi, which allows us to trace platform development from the early phase onward. The three prefectures exhibit broadly similar movements early in the sample, followed by divergence and partial re-convergence later on. Elasticities related to both users and vacancies remain stable and similar in magnitude across regions, suggesting limited geographic heterogeneity in matching responsiveness on the platform.}

This study lays the foundation for understanding the economic mechanisms governing spot gig work, extending labor economics research beyond traditional and even broader gig economy studies. By shedding light on the distinct characteristics of spot work platforms and their efficiency in labor market matching, we offer novel insights into a rapidly evolving segment of the labor market.

\subsection{Related literature}

First, this paper contributes to the empirical literature on estimating the matching function, a fundamental component in macroeconomic models. We examine the trend of matching efficiency in Japanese labor markets via an online spot work matching platform nonparametrically using a novel approach \citep{lange2020beyond}, which shows how to nonparametrically identify the matching function and estimate the matching function allowing for unobserved matching efficacy, without imposing the usual independence assumption between matching efficiency and search on either side of the labor market, allowing for multiple types of jobseekers.
\cite{lange2020beyond} highlight positive correlations between efficiency and market structure such as tightness and so on, which induces a positive bias in the estimates of the vacancy elasticity whenever unobserved matching efficacy is not controlled for, as is the case in the traditional Cobb-Douglas matching function with constant elasticity parameters.\footnote{Using administrative data of users and vacancies, \cite{petrongolo2001looking} summarize the early aggregate studies based on a Cobb-Douglas matching function with the flow of hires on the left-hand side and the stock of unemployment and job vacancies on the right-hand side. In short, the match elasticity with respect to unemployment is in the range 0.5–0.7. In the context of Japanese labor markets, \cite{otani2024nonparametric} uses the nonparametric approach and updates the existing findings reported in \cite{kano2005estimating}, \cite{kambayashi2006vacancy}, \cite{sasaki2008matching}, and \cite{higashi2018spatial} which use the traditional Cobb-Douglas matching function with geographical and occupational category fixed effects to capture geographical and occupational heterogeneity, which are also useful for comparison with other countries' results reported in \cite{bernstein2022matching} and \cite{petrongolo2001looking}.} 

In the context of Japanese labor markets, \cite{otani2024nonparametric} estimates matching efficiency and mismatch in the off-the-job search administrative worker-vacancy-matching platform via Public Employment Security Offices, known as Hello Work. The paper finds a declining trend in matching efficiency, consistent with decreasing job and worker finding rates. 
The implied match elasticity with respect to unemployment is 0.5-0.9, whereas the implied match elasticity with respect to vacancies is between -0.4 and 0.4.
Applying the same approach to proprietary data, \cite{otani2025onthejob} estimates matching efficiency and elasticity in the on-the-job search scouting platform for high-skill employed workers, operated by a private firm in Japan, and compares these with those for Hello Work full-time workers.
This paper complements the above findings.
A series of papers overview the Japanese labor markets in the 2010s and 2020s and provide empirical evidence on matching efficiency, elasticity, and mismatch.

Second, our paper contributes to a nascent literature on alternative work arrangements in labor economics \citep{mas2020alternative}, which are closely related to spot work markets.
The key difference between spot work and classical work is the frequency and flexibility of labor supply and demand.
This paper is the first to estimate matching efficiency and elasticity to the number of registered workers and spot vacancies on the online spot work platform highlighting the key difference from the public part-time worker-job matching platform.
Alternative work arrangements based on spot platforms are growing rapidly in labor markets.
\cite{katz2019rise} focus on alternative work arrangements defined as temporary help agency workers, on-call workers, contract workers, and independent contractors in the United States from 2005 to 2015.
They provide evidence of an increase in these workers but online platform workers, such as ``Uber" and ``Task Rabbit" represented only 0.5\% of all workers in 2015.
By contrast, \cite{kassi2018online} show that the demand for online labor markets increased by 21\% from 2016 to 2018.
\footnote{Utilizing the unique features, empirical literature characterizes the relationship between labor supply and compensation scheme \citep{chen2019value, angrist2021uber, hall2021pricing, butschek2022motivating} and gender wage gap \citep{cullen2018gender, cook2021gender, adams2025gender}.
\cite{angrist2021uber} have the randomized experiment using the ``Uber" service to compare the value of the proportional compensation scheme offered by ride-share companies with taxi compensation.
\cite{butschek2022motivating} implement a field experiment in the gig economy to confirm workers' intrinsic motivation affects the relationship between performance and the compensation scheme. }

Focusing on the important feature of search and matching environment, some transportation industries such as taxi \citep{frechette2019frictions, buchholz2022spatial, lehe2022taxi}, ride-sharing like Uber \citep{guda2019your, rosaia2020competing, castillo2023benefits, castillo2024matching}, bulk shipping \citep{brancaccio2020geography,brancaccio2023search} are well studied.
These studies describe models to explain why inefficiencies occur in the search and matching of sellers and buyers.
% \cite{frechette2019frictions} estimates a dynamic equilibrium model of the New York City taxicab market to assess the importance of regulatory entry restrictions and of matching frictions.
% \cite{rosaia2020competing} investigate a model of competing oligopolistic platforms in the transport market and he find that it would reduce the average number of idle vehicles by 30\%, improving efficiency and reducing vehicle traffic by 12\% in the case of a merger platforms.
% \cite{castillo2024matching} analysis matching failure in which high demand sets off a harmful feedback cycle of few idle drivers, high pickup times, and low earnings, drastically reducing welfare.
However, to our knowledge, there is less literature about the overall features in the spot labor markets where the labor force on customer service, sales, and cleaning service is demanded and this is the first article to investigate the matching efficiency in these markets.

Third, this paper contributes to the growing literature on online job search platforms. The analysis of job matching within actual market institutions has gained prominence with the increasing availability of data from job advertising platforms \citep{autor2019studies}.\footnote{For example, studies such as \cite{kuhn2004internet}, \cite{kuhn2014internet}, and \cite{kroft2014does} use data focused solely on worker status, while papers like \cite{kuhn2013gender}, \cite{hershbein2018recessions}, \cite{brown2016boarding}, and \cite{azar2020concentration} focus exclusively on vacancy data. Other studies, such as \cite{banfi2019high}, \cite{marinescu2018mismatch}, \cite{marinescu2020opening}, and \cite{azar2022estimating} utilize data that captures both worker and vacancy information.}  Much of the literature examines application-level or vacancy-level behavior to capture search behavior and wage elasticity. For instance, \cite{faberman2019intensity} use proprietary application-level data from an online job search engine to investigate the relationship between search intensity and search duration, focusing on lower-skill, hourly jobs for both employed and unemployed workers. Similarly, \cite{kambayashi2025decomposing} estimate elasticities of application, interview attendance, and offer acceptance with respect to posted wages using detailed process-level data for high-skill workers and firms on private job search platforms in Japan. In contrast, this paper takes a broader macro-level view, evaluating the efficiency of the private matching platform itself. To our knowledge, this is the first paper to estimate matching efficiency and elasticity in an online spot work matching platform, providing relatively longer-term insights into private online job search trends and complementing the micro-level studies mentioned above.

\textcolor{black}{
The rest of the paper is organized as follows. 
Section \ref{sec:data} outlines the platform and the Hello Work data. 
Section \ref{sec:model} introduces our model of nonparametric aggregate matching function.
Section \ref{sec:estimation} describes our method of estimation. 
Section \ref{sec:results} reports the main results. 
Section \ref{sec:conclusion} concludes.
}

\section{Data} \label{sec:data}

\subsection{Data source}

First, we use the Report on Employment Service (\textit{Shokugyo Antei Gyomu Tokei}) for the month-level aggregate data from December 2019 to December 2023 to capture trends in matches between unemployed workers and vacancies via the public employment service. 
These datasets include the number of job openings, job seekers, and successful job placements, primarily sourced from the Ministry of Health, Labour and Welfare (MHLW) of Japan, which publishes monthly reports and statistical data on the Public Employment Security Office, commonly known as Hello Work. \textcolor{black}{These series are not seasonally adjusted, because our objective is to estimate the matching function flexibly in a nonparametric way rather than impose seasonal effects through fixed effects.}
Hello Work is a government-operated institution in Japan that provides job seekers with employment counseling, job placement services, and vocational training, playing a critical role in Japan's labor market. 
The data is often used as in \cite{kano2005estimating}, \cite{kambayashi2006vacancy}, \cite{sasaki2008matching}, \cite{kambayashi2013role}, \cite{higashi2018spatial}, and \cite{kawata2021first}. In particular, \cite{shibata2020labor} estimates the traditional Cobb-Douglas matching function, whereas \cite{otani2024nonparametric} estimates the nonparametric matching function.
In this study, we focus on part-time workers in Hello Work for a reasonable comparison.
The sample period is chosen to match the timeframe of the platform data below.

Second, we use proprietary data from Timee, a private company that operates a spot-worker matching platform in Japan, to analyze trends in matches between part-time spot workers and vacancies. Unlike Hello Work, the platform operates as an on-demand staffing service designed to connect businesses with temporary workers for short-term jobs. Its primary users are workers seeking flexible, task-based employment rather than long-term positions, including unemployed individuals and those who want to work on demand. Workers can register on the app for free and immediately access job postings from various companies, while the platform streamlines the matching process for shifts across industries such as food, retail, and logistics.\footnote{Using the same data, \cite{kanayama2026justminimumwagehikes} study the effect of the minimum wage on spot employment on the platform using a Difference-in-Differences approach. See \cite{kanayama2026justminimumwagehikes} for more details about the platform and vacancy-side variables. In addition, \cite{sekiya2026concentration} conduct a field experiment to investigate the effect of their proposed algorithm on job-favoriting and application behavior.} 

The platform's business model allows workers to browse and accept jobs without the need for a traditional hiring process. Instead of charging a subscription fee, the platform earns revenue through fees paid by the businesses utilizing the service. This arrangement benefits both companies and workers by offering flexible, short-term employment opportunities while avoiding the formalities and commitments of long-term contracts. The platform's simplicity and immediacy are its main draws for workers looking to fit employment around their schedules.

Several remarks are noteworthy. First, active job seekers on the platform are defined as registered workers who have activated the Timee app within a given month. \textcolor{black}{Some of these active users may only complete registration in that month without engaging in actual job search activity. Therefore, active users on Timee do not correspond exactly to unemployed job seekers in the Hello Work data.}\footnote{\textcolor{black}{In the Hello Work system, job-seeker registration remains valid, in principle, until the end of the second month following the month of registration, and it can be extended on a monthly basis when the job seeker receives employment services or engages in job search activity.}} Second, unlike Hello Work, where job seekers must enter part-time contracts for a set period and then leave the platform, active job seekers on Timee can match with multiple spot vacancies in a month. Third, some registered workers may use the platform for spot work as a side job. Because our analysis focuses on within-platform search and matching, we do not distinguish such users separately in the analysis.

Panel (b) in Figure \ref{fg:labor_force} shows a diverse range of job categories, with Entertainment, Food Industry, and Office Work occupying significant shares. Other categories like Event Staff, Light Work, and Professional roles also contribute to the platform's job distribution, albeit to a lesser extent. This surge suggests a growing reliance on the spot work platform for flexible employment opportunities, particularly in sectors like Entertainment and Food Industry. The steady rise across multiple categories implies an expanding spot labor market, reflecting a broadening acceptance and utilization of on-demand work arrangements in Japan.

\textcolor{black}{
\subsection{Age and gender comparison}
}

\begin{figure}[!ht]
  \begin{center}
  \subfloat[Age cohort comparison, 2023]{\includegraphics[width = 0.6\textwidth]{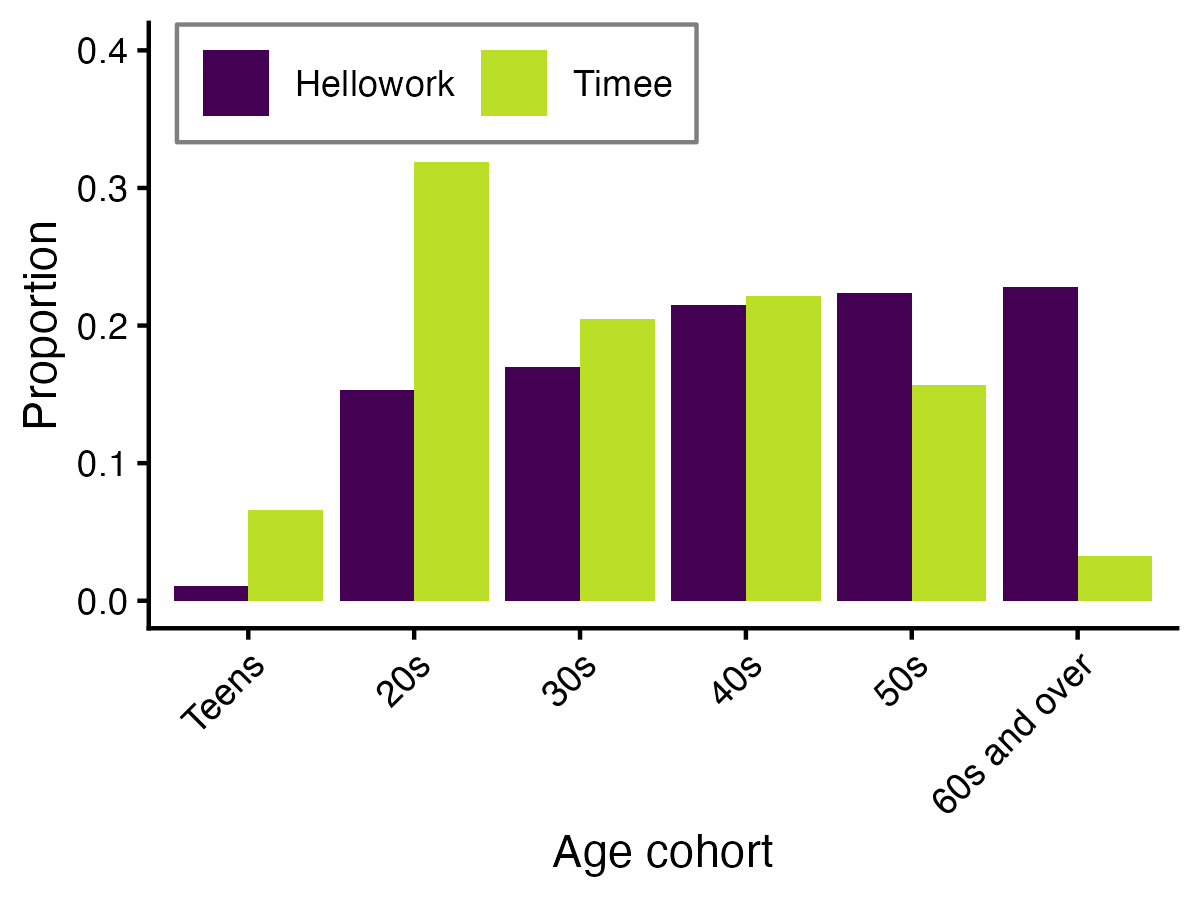}} \\
  \subfloat[Gender comparison, 2023 ]{\includegraphics[width = 0.6\textwidth]{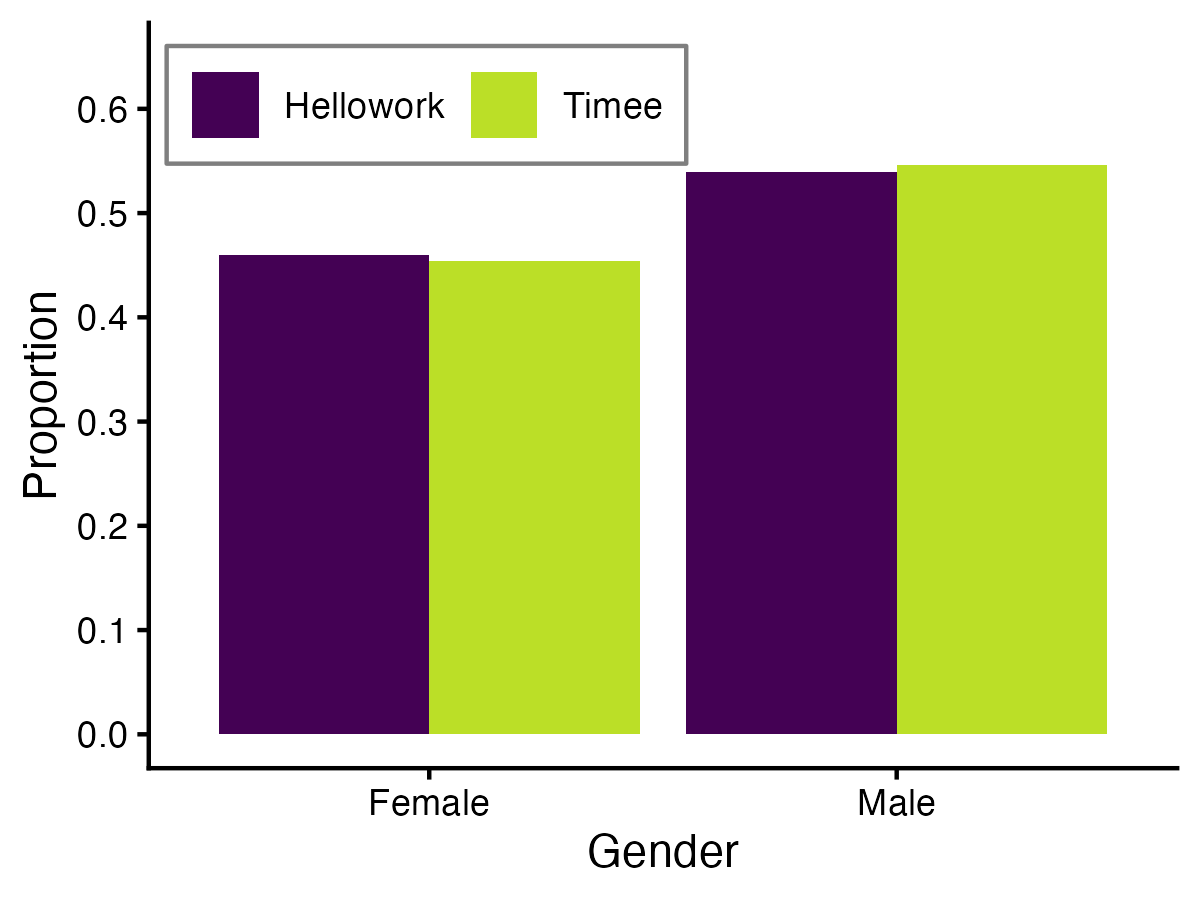}}
  \caption{Age cohort and gender comparison between Hello Work and the platform}
  \label{fg:age_cohort_gender_timee_hellowork}
  \end{center}
  \footnotesize
  Note: Panel (a) shows the age-cohort distribution of hires in 2023. Panel (b) shows the gender distribution of hires in 2023.
\end{figure} 

To explore the difference between Hello Work and the private spot work platform, we compare user characteristics across the two platforms. This comparison helps illustrate qualitative differences in the nature of search and matching across the two platforms.

\textcolor{black}{
Figure \ref{fg:age_cohort_gender_timee_hellowork} presents the age-cohort and gender composition of hires across the two platforms in 2023.
Panel (a) of Figure \ref{fg:age_cohort_gender_timee_hellowork} presents the age-cohort distribution of hires in 2023. The purple bars denote the hiring shares on Hello Work, while the green bars denote those on the private platform. 
The age distribution of hires differs markedly between the two platforms. 
On the private platform, workers in their twenties account for the largest share of hires, at around 30 percent, whereas workers in their fifties and those aged 60 or above account for relatively smaller shares. 
By contrast, on the public platform, workers in their twenties and thirties represent relatively smaller shares, while workers aged 40 and above each account for more than 20 percent of hires in a relatively stable manner. 
These patterns suggest that the two platforms serve different segments of the labor market: the private platform is more strongly oriented toward younger workers, whereas the public platform covers a larger share of middle-aged and older workers.
}

\textcolor{black}{
Panel (b) of Figure \ref{fg:age_cohort_gender_timee_hellowork} presents the distribution of hires by gender in 2023. 
The purple bars denote the hiring shares on Hello Work, while the green bars denote those on the private platform. 
In contrast to the age distribution, the gender composition of hires is broadly similar across the two platforms.
In both the private and public platforms, male hires account for more than 50 percent of total hires, whereas female hires account for less than 50 percent. 
This pattern suggests that the two platforms do not differ substantially in terms of the gender composition of their users.
}

\subsection{Trend comparison}

\begin{figure}[!ht]
  \begin{center}
  \subfloat[User $U$, vacancy $V$, and tightness ($\frac{V}{U}$)]{\includegraphics[width = 0.37\textwidth]
  {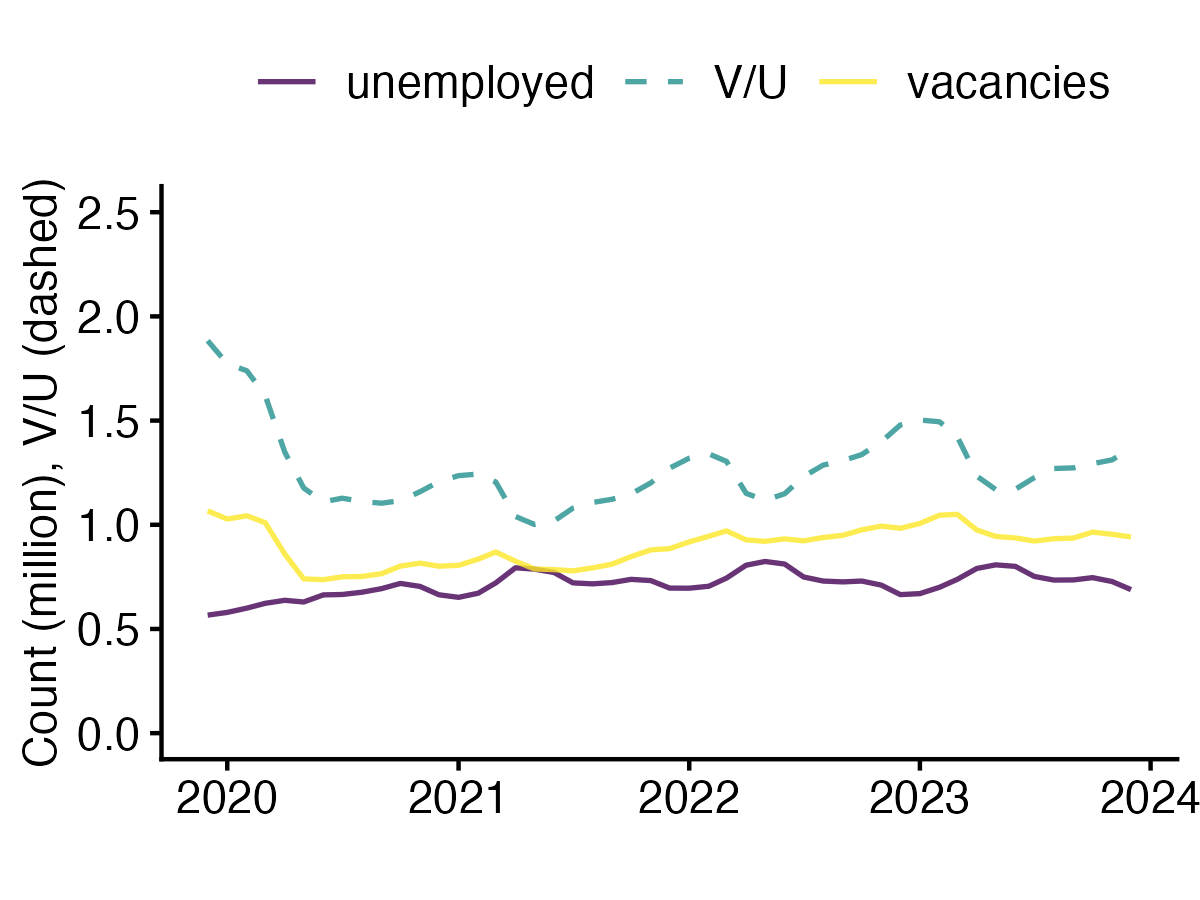}\includegraphics[width = 0.37\textwidth]
  {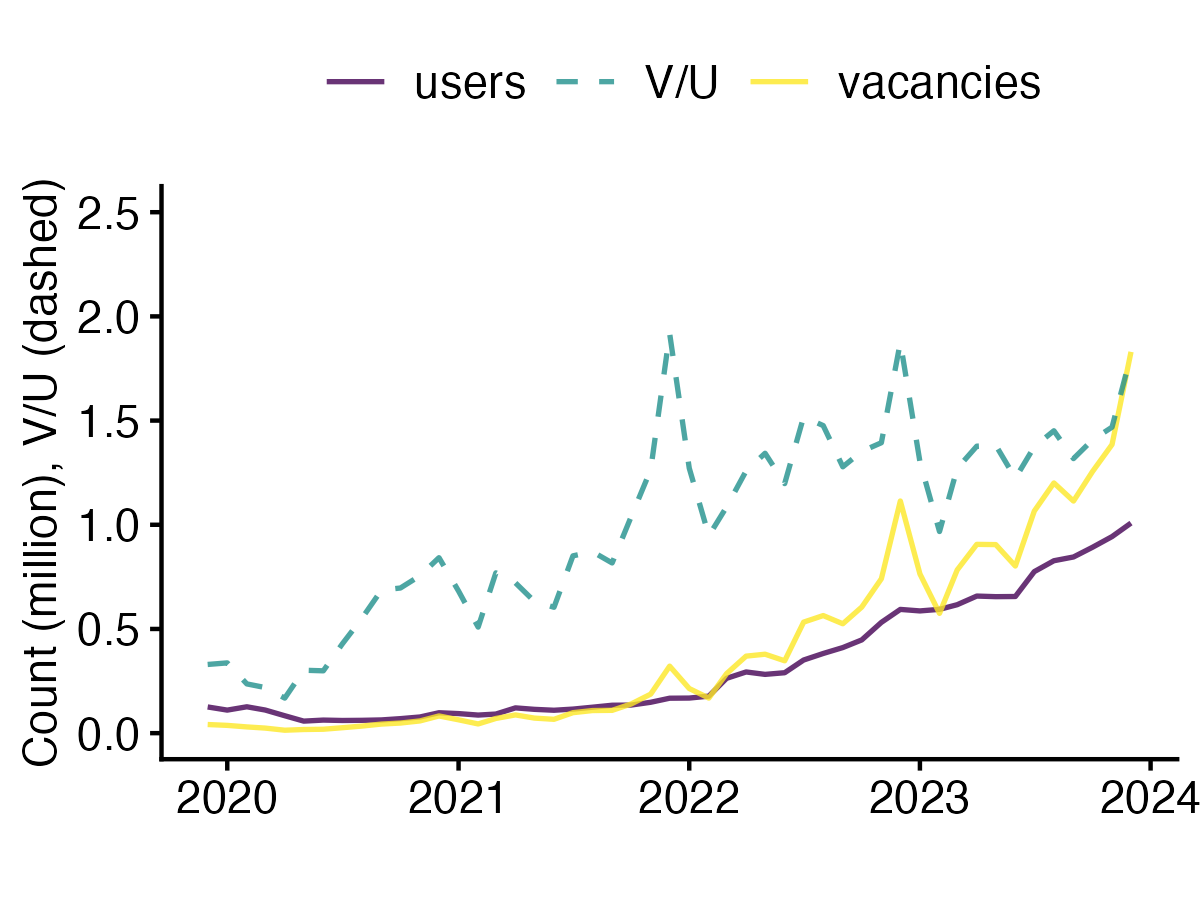}}\\
  \subfloat[Hire $H$]{\includegraphics[width = 0.37\textwidth]
  {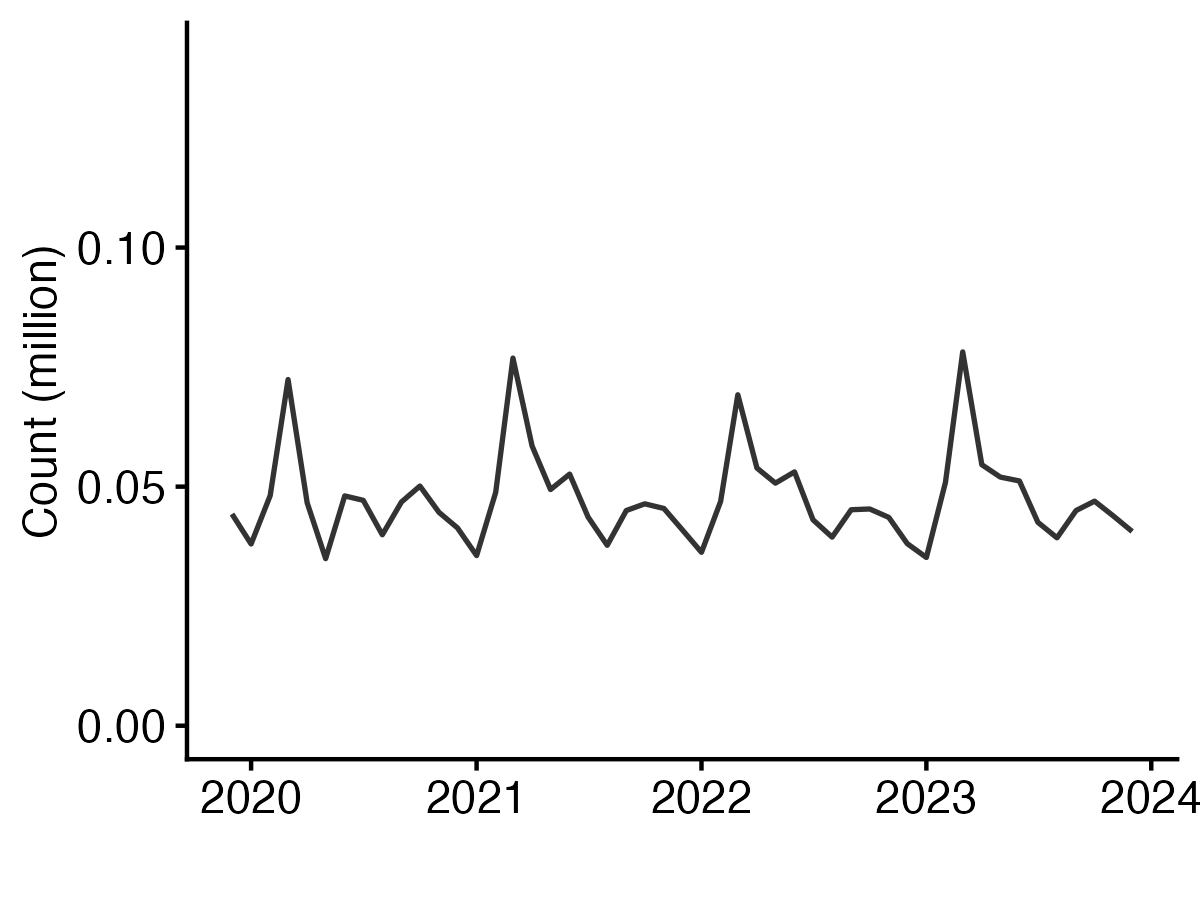}\includegraphics[width = 0.37\textwidth]
  {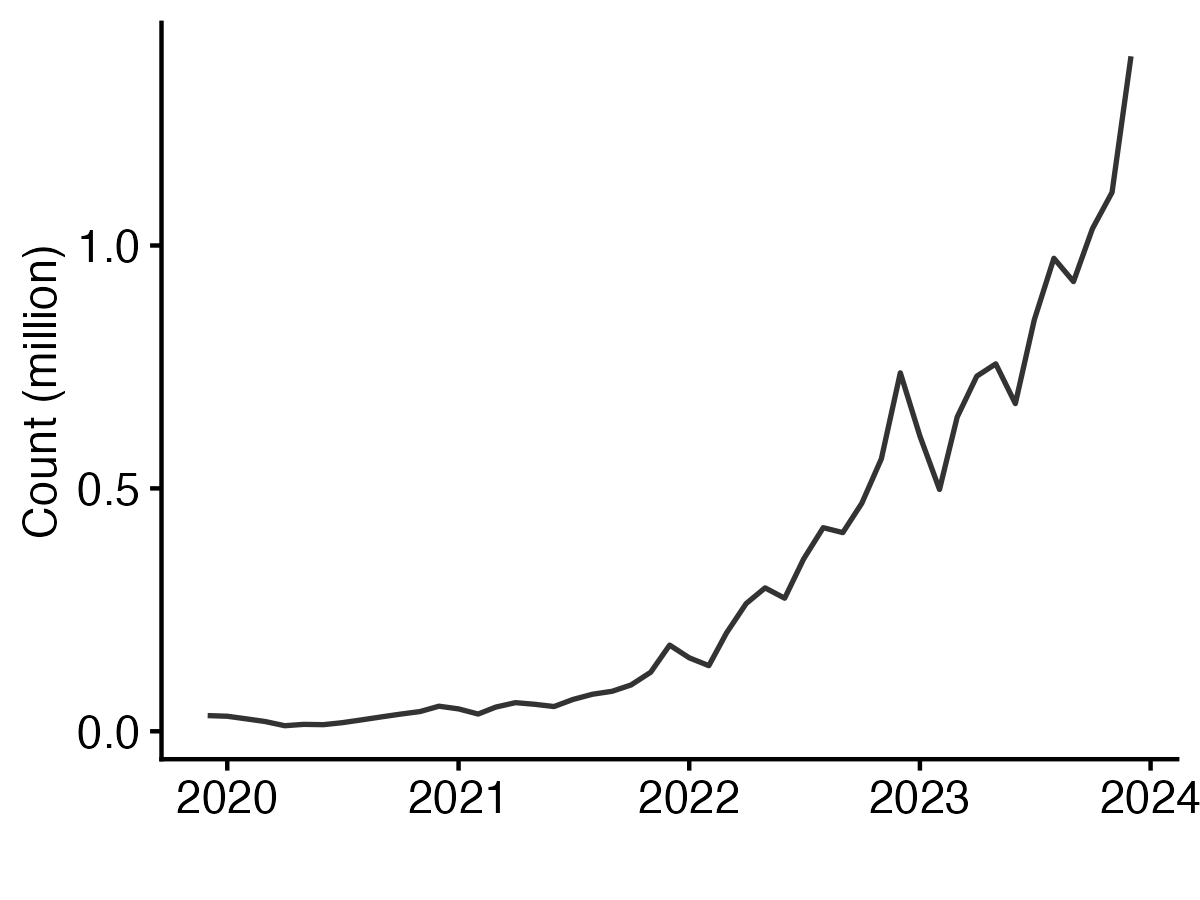}}
  \\
  \subfloat[Job and worker finding rate ($\frac{H}{U}$,$\frac{H}{V}$)]{\includegraphics[width = 0.37\textwidth]
  {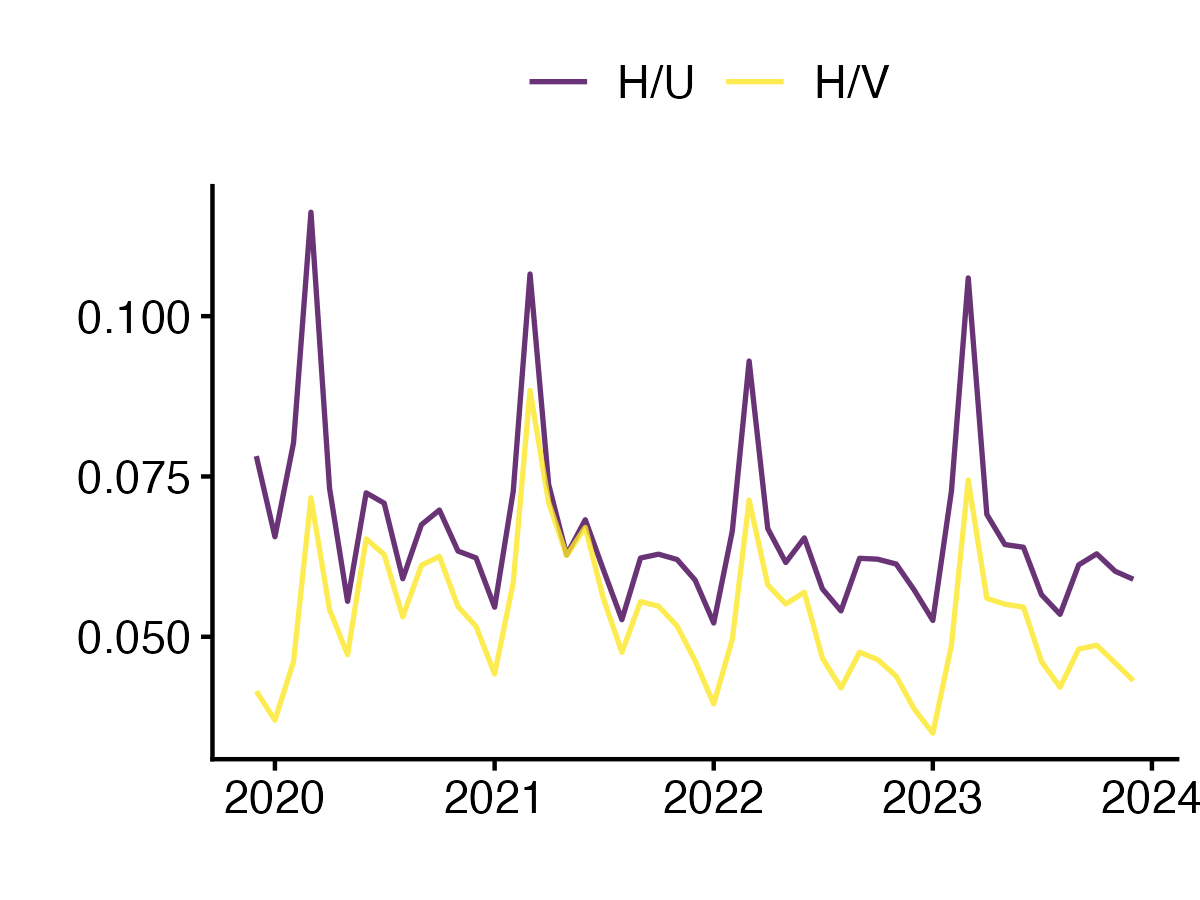}\includegraphics[width = 0.37\textwidth]
  {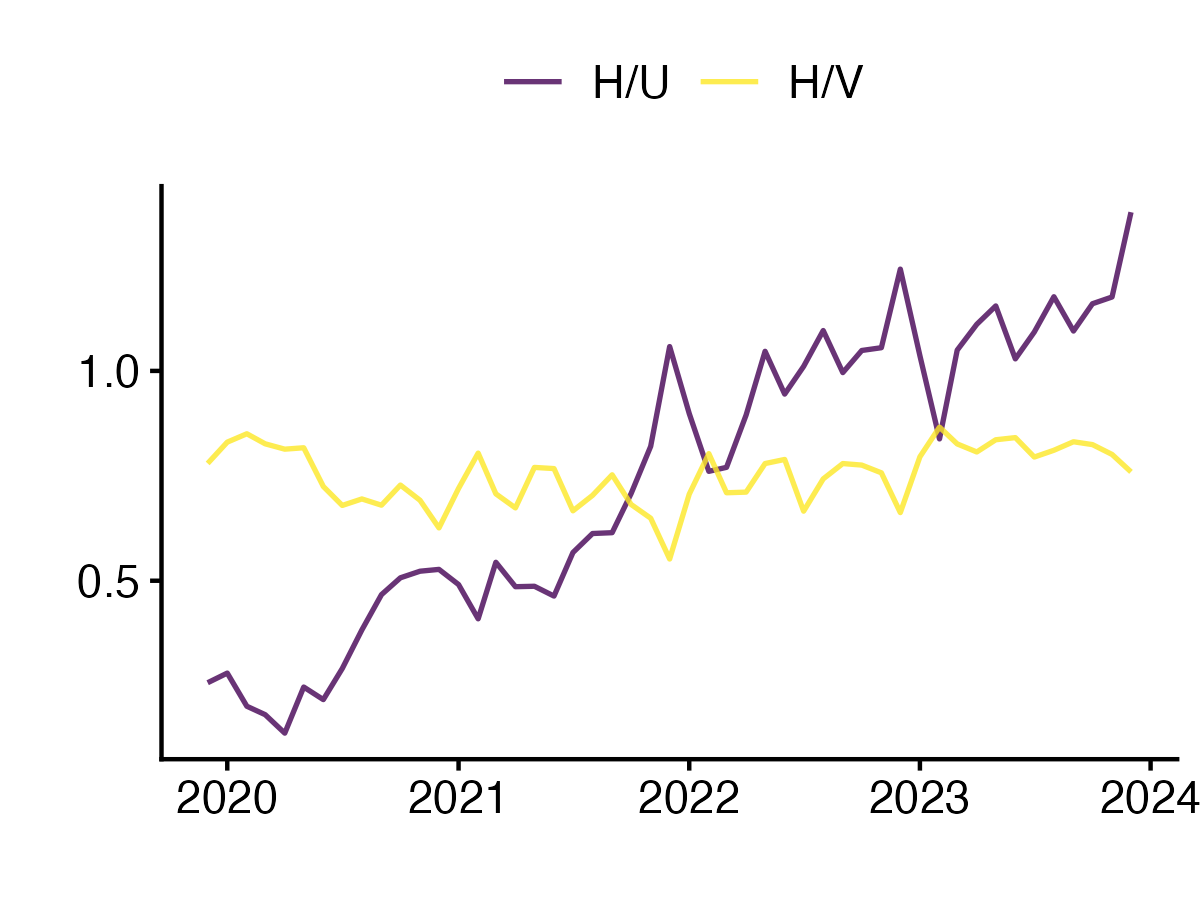}}
  \caption{Trends of key variables: Hello Work part-time (left) vs platform (right) 2019-2023}
  \label{fg:unemployed_vacancy_month_aggregate} 
  \end{center}
  \footnotesize
  %Note: 
\end{figure} 

Figure \ref{fg:unemployed_vacancy_month_aggregate} provides a comparative analysis of labor market dynamics between Hello Work (left panel) and the private spot work platform (right panel) from December 2019 to December 2023.

\textcolor{black}{
First, we discuss the left panel of Hello Work for part-time workers and jobs. In panel (a), the number of unemployed individuals remains broadly stable over the observed period, hovering around 0.8 million. Vacancies and labor market tightness ($V/U$) decline in the first half of 2020, likely reflecting the disruption caused by the COVID-19 pandemic, but remain broadly stable thereafter without a clear recovery to the pre-pandemic level.}\footnote{\textcolor{black}{The decline in vacancies and labor market tightness in the first half of 2020 is consistent with the substantial disruption to labor demand caused by the COVID-19 pandemic in Japan \citep{fukui2020job, kawaguchi2021can, kikuchi2021suffers, fukai2021describing, ando2023discontinuous, higashi2025did}. This period coincided with a sharp deterioration in economic activity, particularly in face-to-face service sectors where part-time employment is prevalent.}} Panel (b) shows that the hiring count through Hello Work remains low, with a small oscillating pattern throughout the period. This stable but limited hiring activity suggests that the platform may face constraints or inefficiencies in increasing the job match rate for part-time workers. In panel (c), the job and worker finding rates ($H/U$ and $H/V$) exhibit a gradual decline over time, suggesting reduced effectiveness in matching job seekers with available vacancies via Hello Work. This decline might reflect a growing preference for alternative job search methods such as the spot-work platform.

In contrast, the right panels reflect the dynamics of the private platform, where a distinctive pattern emerges. Unlike Hello Work's part-time flow from unemployed to employed, workers on this platform do not automatically leave after securing a match, which keeps a larger pool of active users. The number of registered users increases gradually, while the number of vacancies rises sharply, especially post-2022, leading to a notable rise in labor market tightness ($V/U$). This reflects an expanding demand for spot employment opportunities. The hiring count in panel (b) shows a rapid upward trajectory starting around 2022, underscoring a marked increase in successful matches.

Panel (c) illustrates the job and worker finding rates ($H/U$ and $H/V$), highlighting significant differences compared to Hello Work. The job finding rate ($H/U$) continues to rise, reaching values close to 1.3, indicating that each worker matches with multiple vacancies per month. On the other hand, the worker finding rate ($H/V$) remains stable around 0.8, suggesting that 80\% of the available jobs are successfully filled. This pattern demonstrates the platform's increasing efficiency in matching workers to vacancies. It indicates a maturing spot labor market that still exhibits strong growth.

\section{Model} \label{sec:model}
\subsection{Nonparametric aggregate matching function}
Our primary focus is on analyzing matching efficiency and matching elasticity in relation to the number of registered workers and available vacancies in the labor market, facilitated by an online spot work matching platform operated by Timee in Japan. A matching function derived from search models is central to labor economics.\footnote{See \cite{pissarides2000equilibrium}, \cite{petrongolo2001looking}, and \cite{rogerson2005search} for further reference.} This function is based on the premise of random search from both sides of the labor market, where job seekers represent labor supply and firms posting vacancies represent labor demand. 

To estimate the matching function and recover matching efficiency, we adopt the novel approach proposed by \cite{lange2020beyond}.\footnote{\cite{lange2020beyond} also integrate search effort \citep{mukoyama2018job} and a recruitment intensity index \citep{davis2013establishment}.} This approach addresses two critical issues: the endogeneity of matching efficiency \citep{borowczyk2013accounting} and the limitations of the Cobb-Douglas specification, which assumes constant matching elasticity. To overcome these challenges, \cite{lange2020beyond} propose a nonparametric identification and estimation framework for matching efficiency, under specific conditions that will be explored later in this paper.

Let $(A, U, V)$ denote random variables while realizations are subscripted by time $t$.
We define that matching function $m_t(\cdot, \cdot)$ maps period-$t$ unemployed workers $U_t$, per-capita search efficacy/matching efficiency of the unemployed workers $A_t$, and vacancies $V_t$ into hires $H_t$.
\textcolor{black}{Let $G(H,U,V)$ and $F(A,U)$ denote the joint distribution of $(H,U,V)$ and $(A,U)$.}
Let us assume that $V$ and $A$ are independent conditional on $U$, that is $ V\mathop{\perp} A| U$.
Let also assume that the matching function $m(AU, V): \mathbb{R}^{2}_{+} \rightarrow \mathbb{R}_{+}$ has constant returns to scale (CRS).
Then, by applying nonparametric identification results of \cite{matzkin2003nonparametric}, Proposition 1 of \cite{lange2020beyond} proves that $G(H, U, V)$ identifies $F(A, U)$ and $m(AU, V):\mathbb{R}^{2}_{+} \rightarrow \mathbb{R}_{+}$ up to a normalization of $A$ at one point denoted as $A_0$ of the support of $(A, U, V)$.

\subsection{Potential Limitation}
The first limitation of our approach is that the matching efficiency term $A$ may reflect endogenous variation in search effort or hiring standards. As noted in previous work, firms may adjust recruiting intensity and acceptance thresholds in response to labor market conditions, while compositional mismatch can also affect aggregate matching rates.\footnote{For example, \cite{sedlavcek2014match} shows that firms' hiring standards can be countercyclical. Similarly, worker search effort \citep{gomme2015worker} and firm recruiting intensity \citep{gavazza2018aggregate} vary over the business cycle. Changes in the composition of workers and jobs can affect the aggregate matching rate, appearing as a change in efficiency \citep{csahin2014mismatch}.} These mechanisms could induce a correlation between $A$ and $V$, violating the first identification assumption. Although the spot-work setting may attenuate such effects—since employers generally do not engage in selection or screening, and job tasks are relatively homogeneous compared to standard full-time jobs, thereby limiting occupational composition variation—we acknowledge this as a potential source of bias.

Another limitation concerns the interpretation of matching efficiency and elasticity in a gig work environment that differs from traditional unemployment-to-employment models. Platform users are not necessarily unemployed and can take multiple jobs within a period without exiting the market. While this does not pose a problem for estimating matching functions per se—as in the Cobb-Douglas matching function on the TaskRabbit gig work platform in \cite{cullen2021outsourcing}—it complicates macroeconomic interpretation of finding rates and elasticities \citep{wolthoff2014s}. Additionally, although we impose constant returns to scale (CRS) as is standard, online platforms may exhibit increasing returns due to network externalities. In such cases, the estimated efficiency term may also capture scale effects, not purely frictions, altering its structural interpretation \citep{petrongolo2001looking}. 

Overall, while the estimation remains valid, care must be taken when using the estimated elasticities and efficiency measures as inputs in macroeconomic models or when interpreting them in the context of standard labor market dynamics.

\section{Estimation} \label{sec:estimation}

Following \cite{lange2020beyond}, we begin by estimating \( F(A_0 | U) \) across the support of \( U \). To achieve this, we use the distribution of hires conditional on users, \( U \), and observed vacancies, \( V \). Specifically, we have:

\[
F(A_0|\psi U_0) = G_{H|U,V}(\psi H_0|\psi U_0, \psi V_0) \quad \text{for any arbitrary scalar } \psi,
\]

\[
F(\psi A_0|\lambda U_0) = G_{H|U,V}(\psi H_0|\lambda U_0, \psi V_0) \quad \text{where } \lambda > 0 \text{ is a scaling factor},
\]
where \( F(A_0|\psi U_0) \) and \( G_{H|U,V} \) represent the respective conditional distributions. By varying the parameters \( (\psi, \lambda) \), we can trace out \( F(A | U) \) across the entire support of \( (A, U) \).

Given that our data is finite, we rely on an estimate of \( G_{H|U,V} \) for the constructive estimator. Consider an arbitrary point \( (H_\tau, U_\tau, V_\tau) \). To obtain \( G(H_\tau | U_\tau, V_\tau) \), we calculate the proportion of observations with fewer hires than \( H_\tau \), taken from observations proximate to \( (U_\tau, V_\tau) \) in the \( (U, V) \)-space. In practice, this is done by averaging across all observations, assigning smaller weights to those with values \( (U_t, V_t) \) distant from \( (U_\tau, V_\tau) \) via a kernel that discounts distant observations. The resulting estimate is expressed as:

\[
F(\psi A_0 | \lambda U_0) = G_{H|U,V}(\psi H_0 | \lambda U_0, \psi V_0),
\]

\[
\hat{F}(\psi A_0 | \lambda U_0) = \sum 1(H_t < \psi H_0) \kappa(U_t, V_t, \lambda U_0, \psi V_0),
\]

where \( \kappa(.) \) denotes a bivariate normal kernel with bandwidth 0.01.

Once the distribution function \( F(A | U) \) is recovered, we invert \( F(A_t | U_t) \) to derive \( A_t \) for all observations in the dataset, using:

\[
A_t = F^{-1}(G(H_t | U_t, V_t) | U_t),
\]

Finally, we recover the matching function as:

\[
m(A_t U_t, V_t) = G^{-1}(F(A_t | U_t) | U_t).
\]

To compute matching elasticities, we employ a linear regression, projecting hires onto the values of vacancies and users, interacted with implied matching efficiency. The resulting estimates approximate the derivatives of the matching function with respect to vacancies and users, interacted with implied matching efficiency. This provides an estimate of the elasticity of the matching function.\footnote{The matching elasticity with respect to users $\frac{d \log m(AU,V)}{d \log U}=\frac{d m(AU,V)}{d U}\frac{U}{H}=\frac{d m(AU,V)}{d AU}\frac{d AU}{dU}\frac{U}{H}=\frac{d m(AU,V)}{d AU}\frac{AU}{H}=\frac{d \log m(AU,V)}{d \log AU}$ is obtained from the regression coefficient of $H$ on $AU$ and multiplying it by $\frac{AU}{H}$. }

\textcolor{black}{Note that the primary objective of this paper is point estimation of matching efficiency and elasticity. Evaluating sampling uncertainty for these estimates is technically challenging in our setting, because the procedure combines nonparametric conditional distribution estimation for dependent time-series data with subsequent inversion and plug-in steps. For this reason, we focus on point estimates and leave a rigorous assessment of sampling error, such as standard errors or error bands, for future research.}

\section{Results} \label{sec:results}

\subsection{Matching efficiency and elasticity in the platform}

\begin{figure}[!ht]
  \begin{center}
  \subfloat[Matching Efficiency ($A$)]{\includegraphics[width = 0.37\textwidth]
  {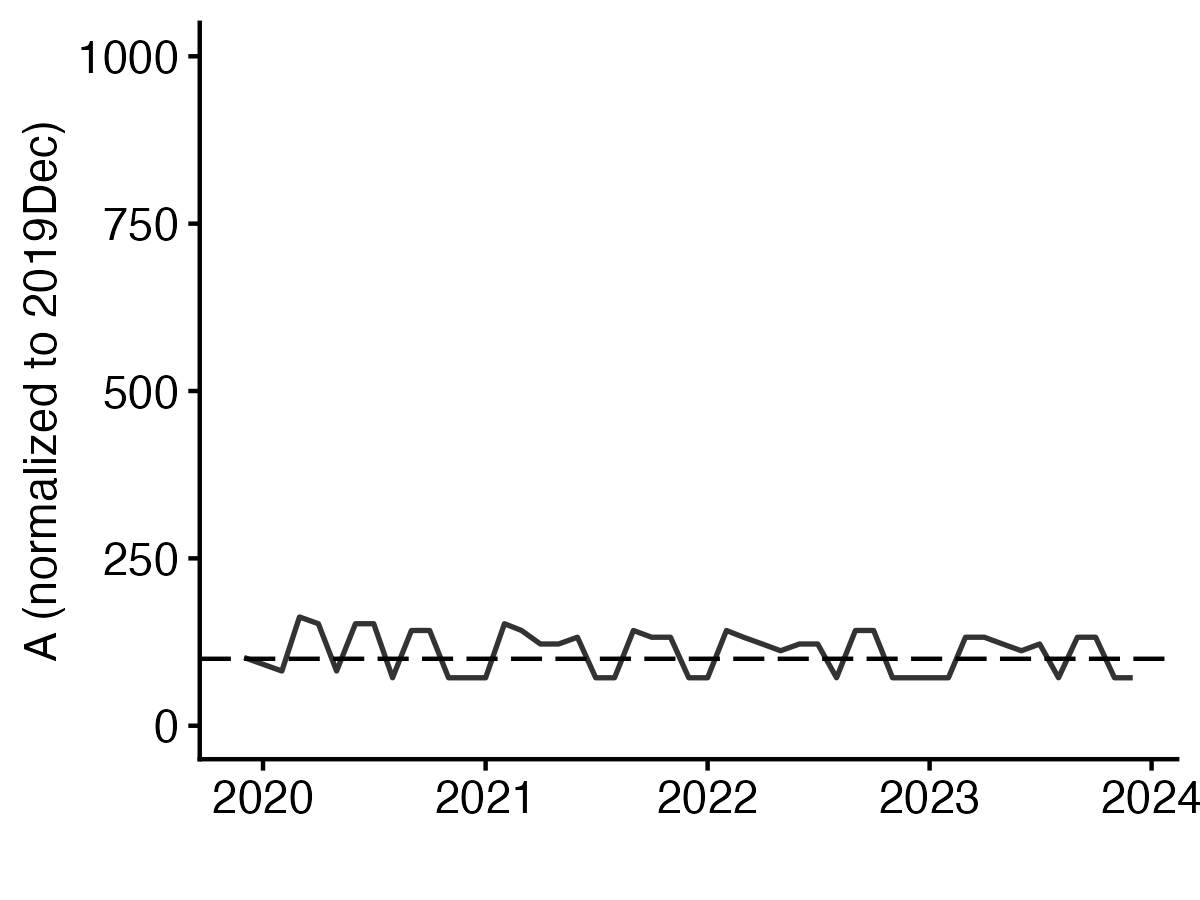}\includegraphics[width = 0.37\textwidth]
  {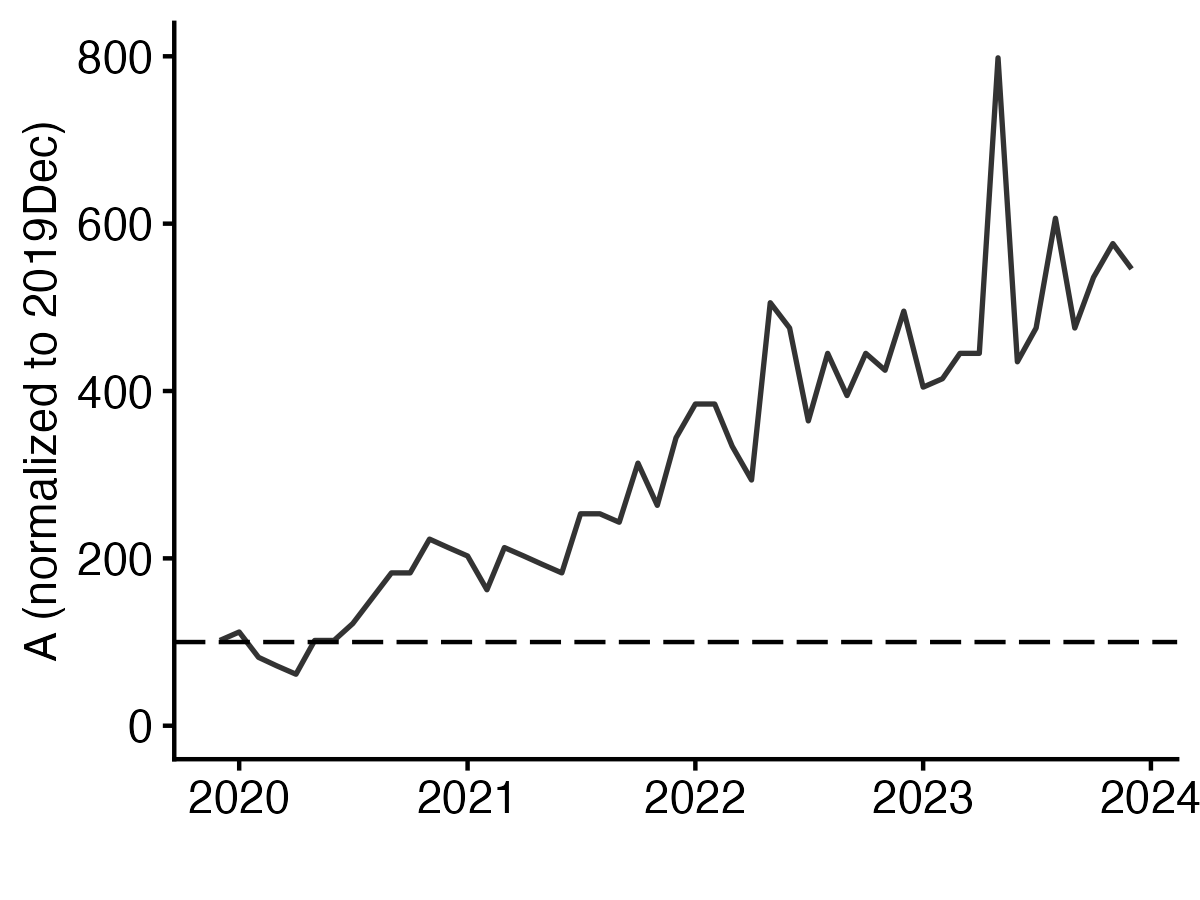}}\\
  \subfloat[Matching Elasticity ($\frac{d\ln m}{d \ln AU}$, $\frac{d\ln m}{d\ln V}$)]{\includegraphics[width = 0.37\textwidth]{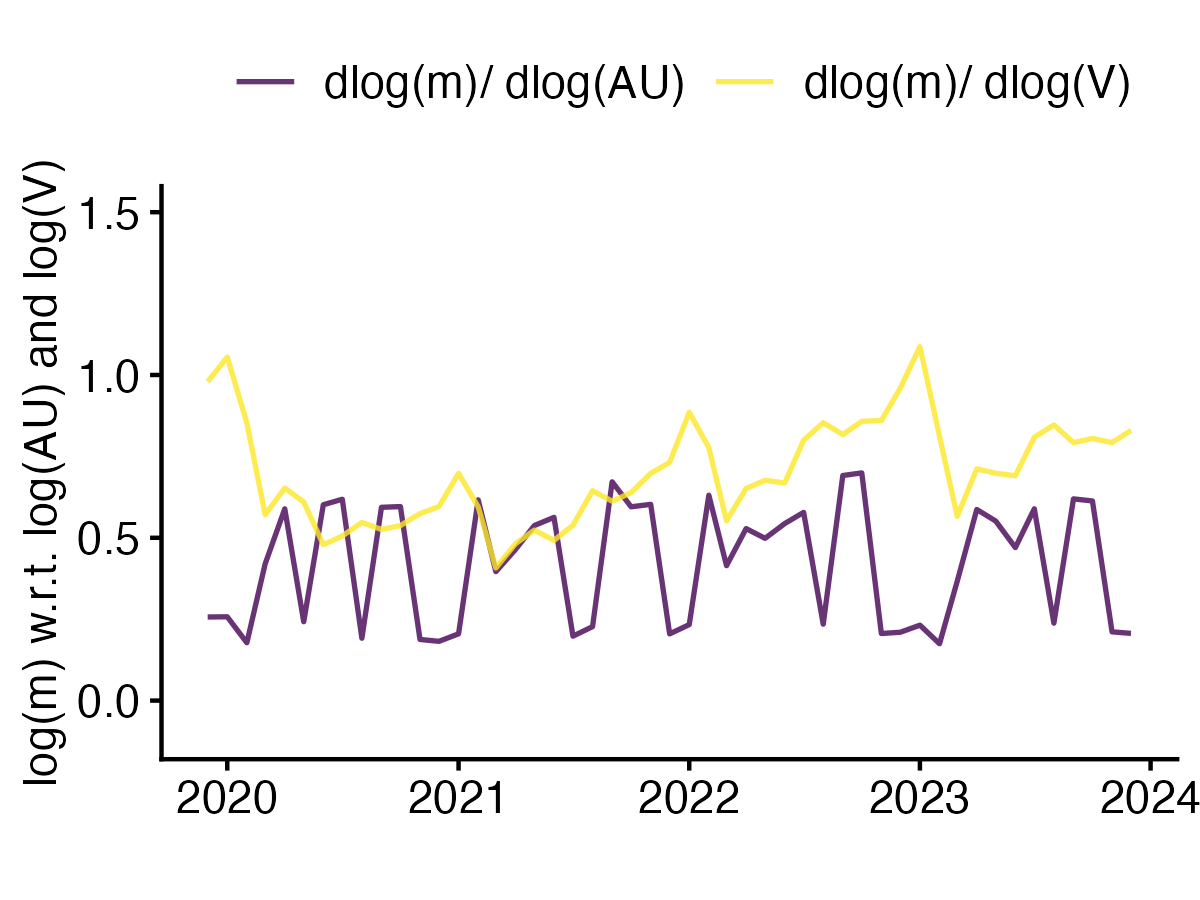}\includegraphics[width = 0.37\textwidth]
  {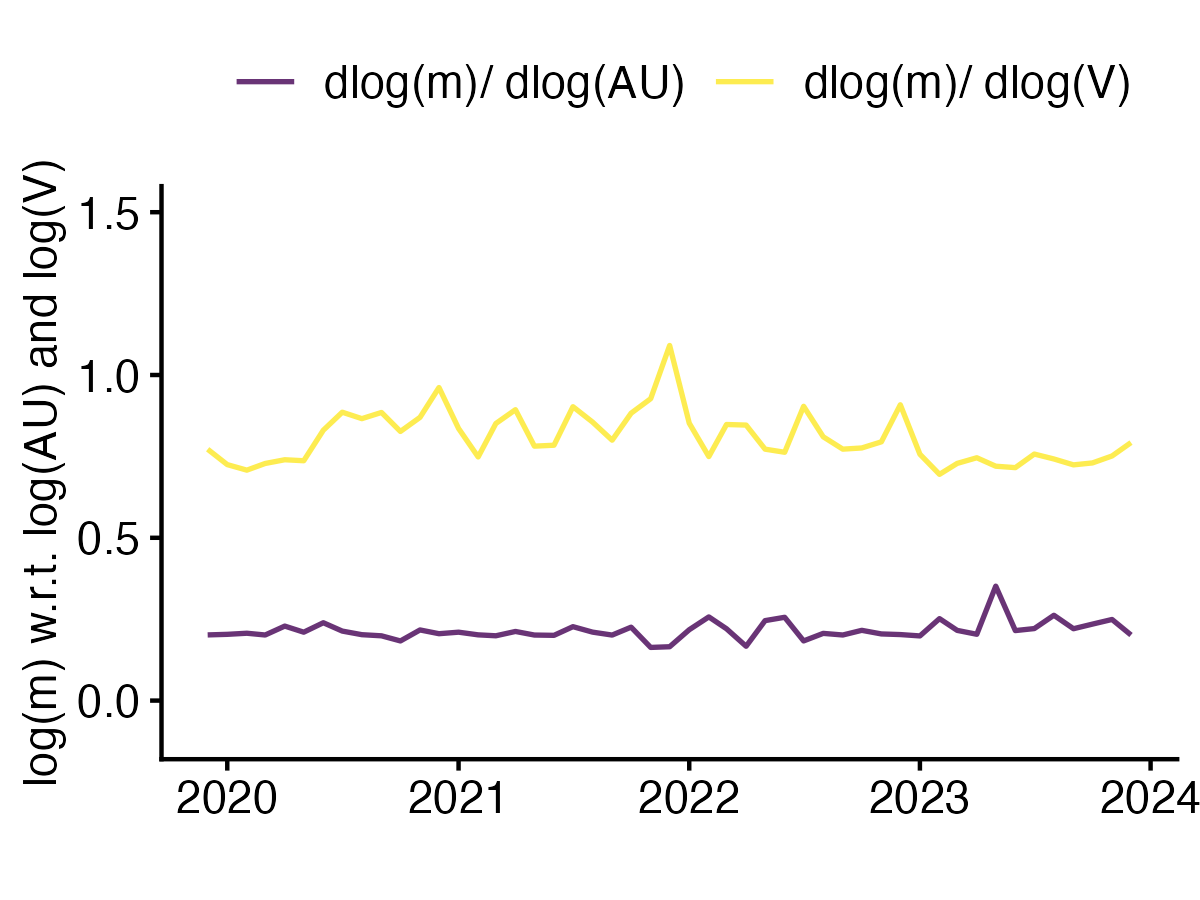}}
  \caption{\textcolor{black}{Hello Work part-time vs platform 2019-2023}}
  \label{fg:matching_efficiency_month_aggregate} 
  \end{center}
  \footnotesize
  Note: For confidentiality reasons, we normalize the efficiency to 2019 December for each platform.
\end{figure} 

Figure \ref{fg:matching_efficiency_month_aggregate} \textcolor{black}{illustrates the trends in matching efficiency, normalized so that December 2019 equals 100,} and matching elasticity for both Hello Work (left panels) and the private platform (right panels) from December 2019 to December 2023.
\textcolor{black}{In panel (a), the dashed horizontal line indicates this baseline level of 100, corresponding to the normalized matching efficiency in December 2019.}
The matching efficiency for Hello Work part-time remains relatively stable around the baseline, fluctuating within a narrow range without any significant upward or downward trends. In contrast, the private platform shows a markedly different pattern, with matching efficiency experiencing sharp increases, peaking around May 2023 at around 800, which implies that the efficiency is about eight times larger than in the initial period in December 2019. 

\textcolor{black}{Panel (b) presents the matching elasticities with respect to users and vacancies for both platforms. For Hello Work (left panel), the elasticity with respect to users fluctuates between approximately 0.3 and 0.5, while the elasticity with respect to vacancies ranges roughly from 0.4 to slightly above 1.0, with somewhat higher volatility, particularly in the earlier years. This suggests that the matching process is only moderately responsive to changes in users and vacancies and exhibits limited variation over time.}\footnote{The estimated elasticities differ from those reported in \cite{otani2024nonparametric}, likely due to the differing lengths of time horizons considered. Specifically, \cite{otani2024nonparametric} includes data from 1972 to 2013, a period covering economic booms and busts, which captures a broader range of labor market dynamics.}

\if0
In contrast, the private platform (right panel) exhibits consistently stable and relatively lower elasticities. The elasticity with respect to users remains relatively stable, ranging between 0.2 and 0.3. The vacancy elasticity is particularly elevated and stable, mostly ranging between 0.7 and 1.1. These patterns indicate that the private platform is more strongly responsive to changes in vacancy posting, even though the user-side responsiveness appears more limited. Overall, the platform demonstrates a matching process that is more vacancy-driven and systematically more efficient in adapting to changes in job demand conditions than the public counterpart.
\fi

In contrast, the private platform (right panel) exhibits consistently stable and relatively lower elasticities. The elasticity with respect to users remains relatively stable, ranging between 0.2 and 0.3. The vacancy elasticity is particularly elevated and stable, mostly ranging between 0.7 and 1.1.
\textcolor{black}{One caveat in interpreting this difference is that, in the platform data, an active user is counted only once within a month even if the same worker matches with multiple jobs, whereas vacancies are recorded at the posting level. This data structure may mechanically attenuate the estimated user elasticity relative to the vacancy elasticity. Nevertheless, even taking this measurement issue into account, the vacancy-side elasticity is substantially larger than the user-side elasticity throughout the sample period. Therefore, the estimates potentially demonstrate that the platform's matching process is more strongly driven by vacancy-side conditions than by active-user inflows.}

\subsection{Prefecture-level matching efficiency and elasticity in the platform}

\begin{figure}[!ht]
  \begin{center}
  \subfloat[User ($U$)]{\includegraphics[width = 0.37\textwidth]
  {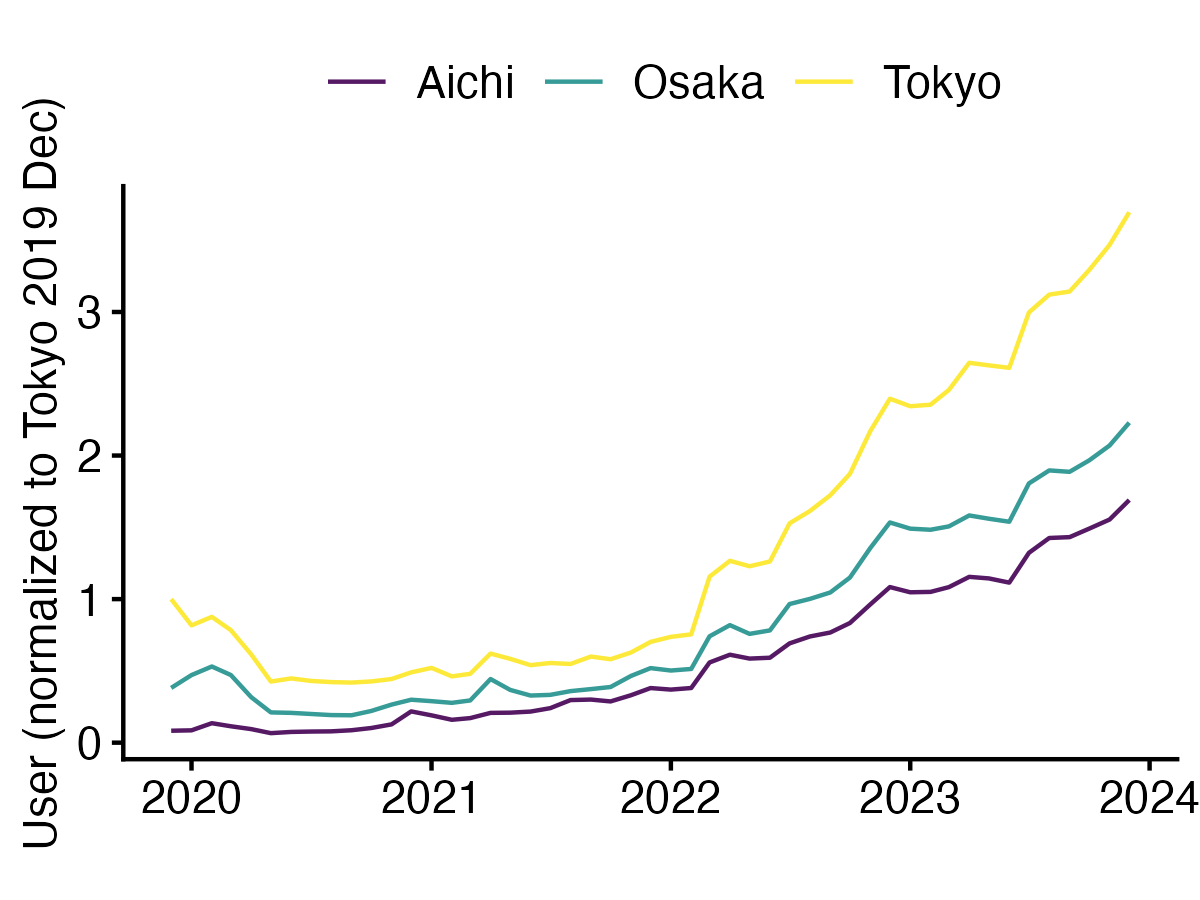}}
  \subfloat[Vacancy ($V$)]{\includegraphics[width = 0.37\textwidth]
  {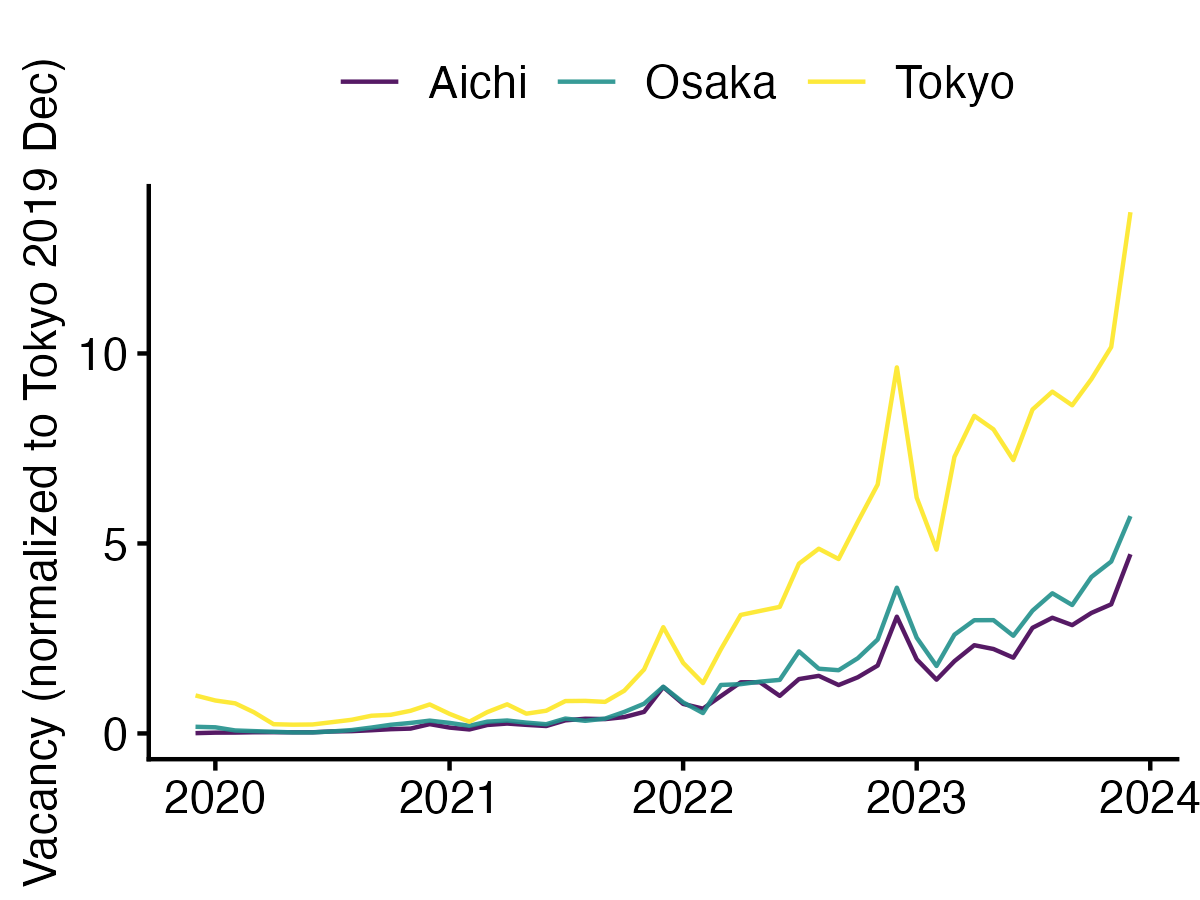}}\\
  \subfloat[Hire ($H$)]{\includegraphics[width = 0.37\textwidth]
  {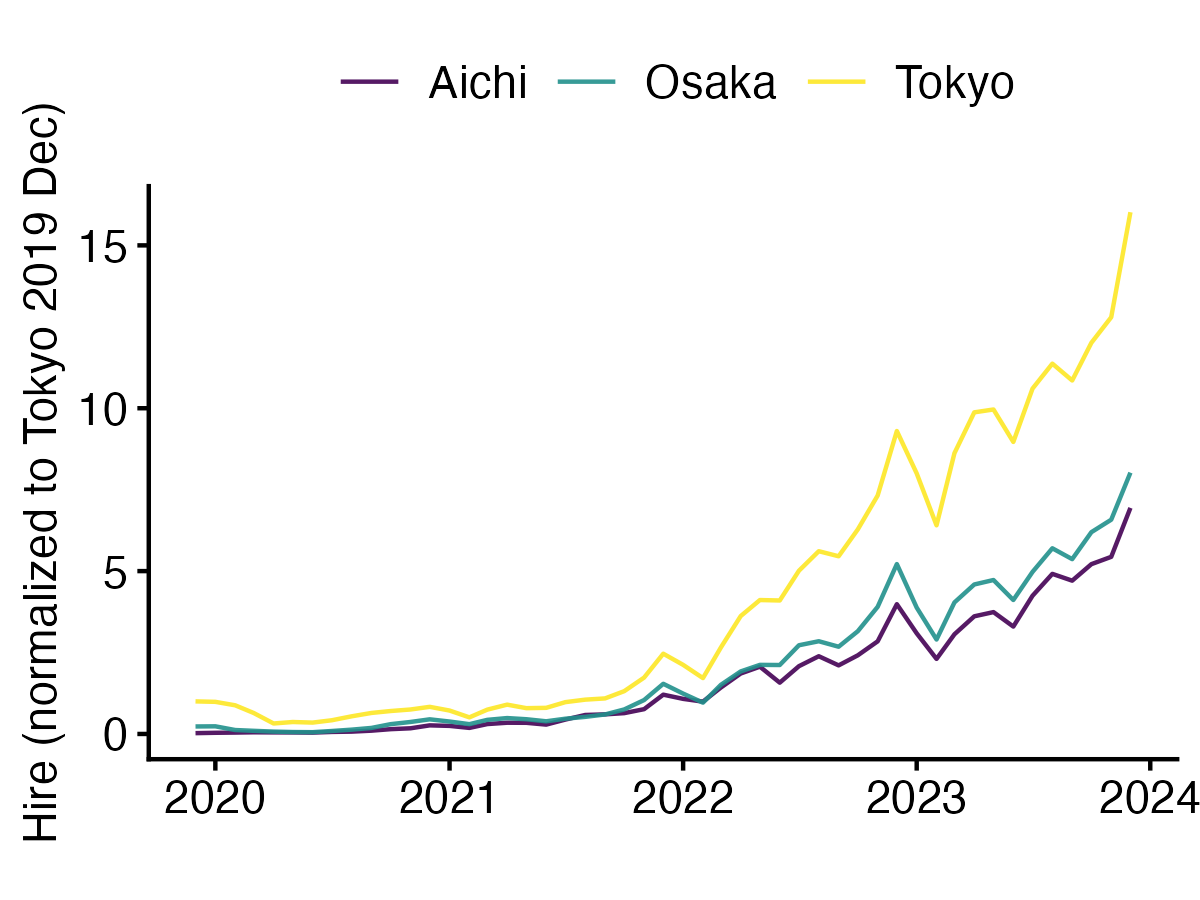}}
  %\subfloat[($V/U$)]{\includegraphics[width = 0.37\textwidth]  {figuretable/matching_function_project/tightness_month_aggregate_platform_each_industry.png}}\\
  \subfloat[Matching Efficiency ($A$)]{\includegraphics[width = 0.37\textwidth]
  {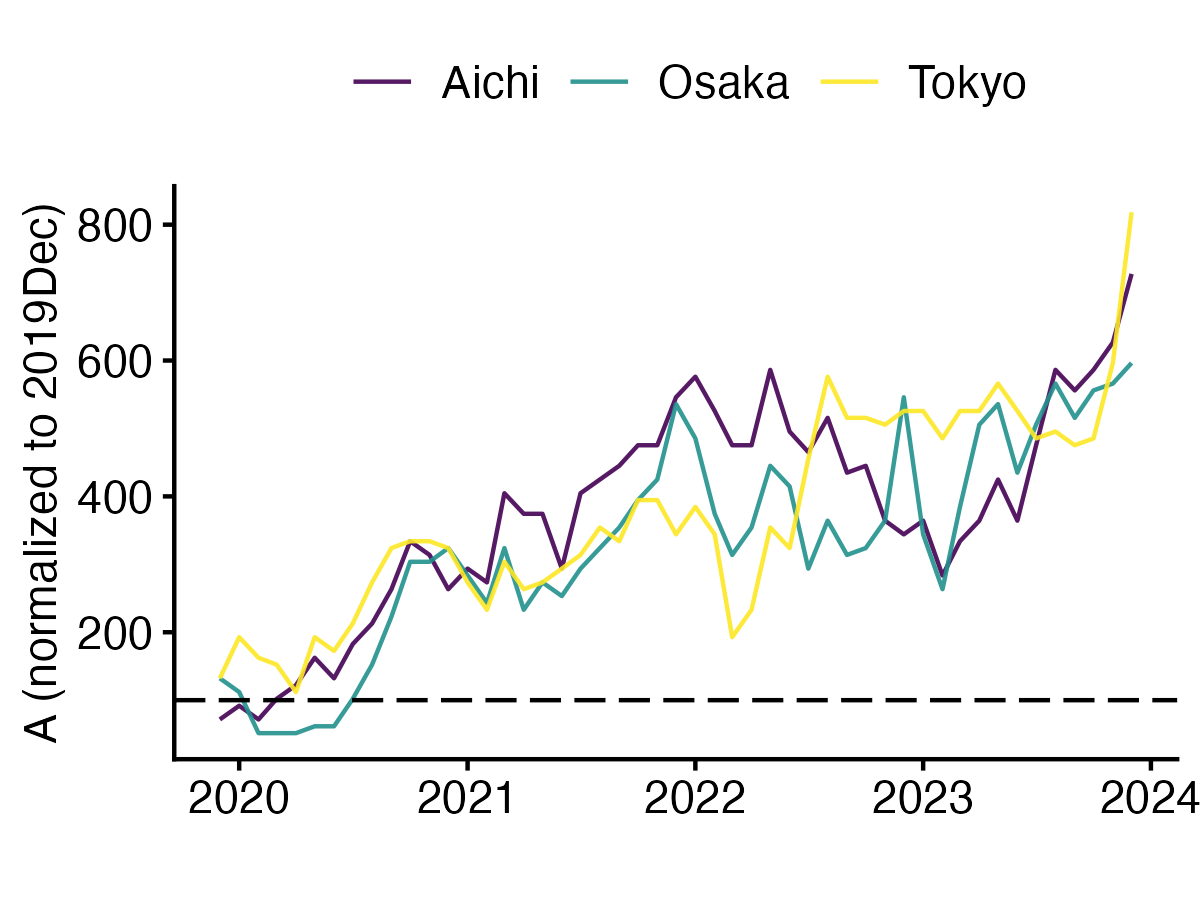}}\\
  \subfloat[Matching Elasticity ($\frac{d\ln m}{d \ln AU}$)]{\includegraphics[width = 0.37\textwidth]
  {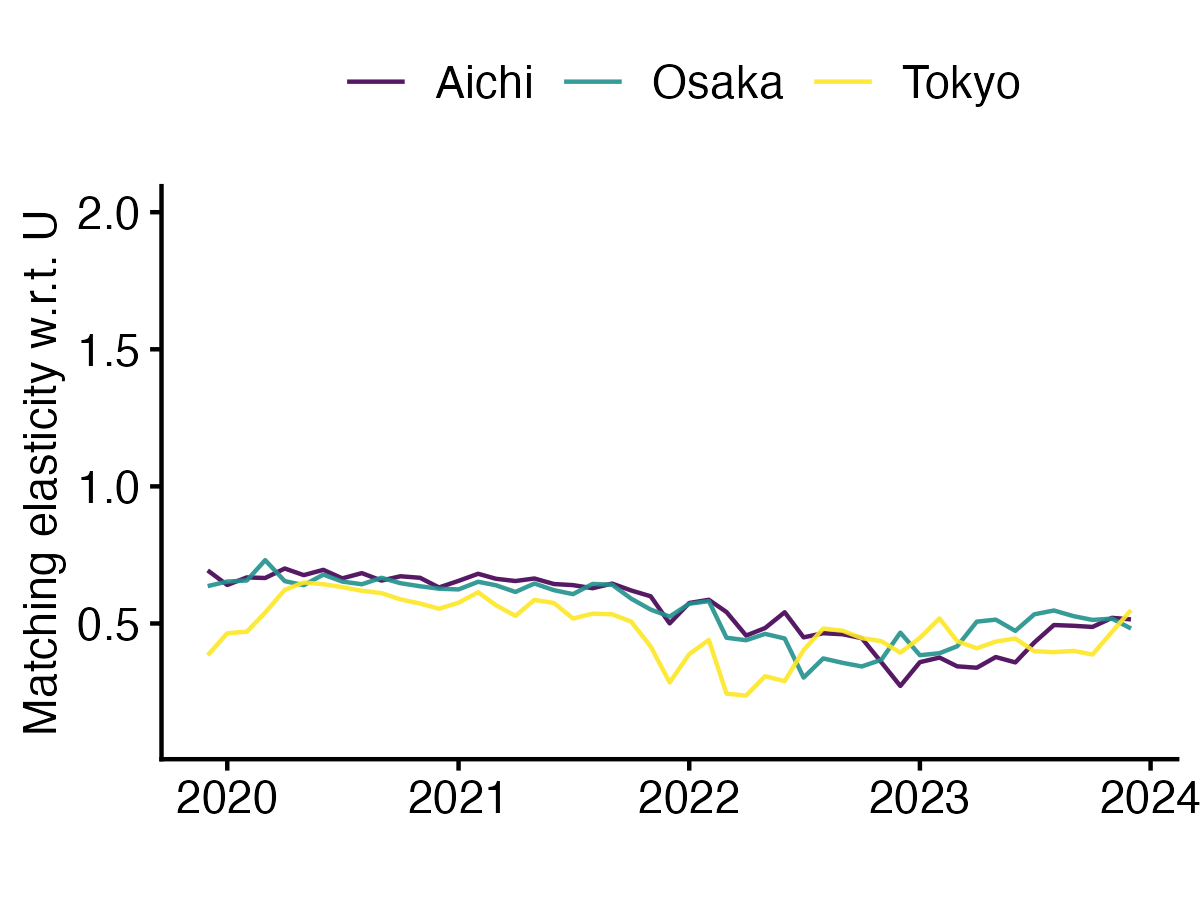}}
  \subfloat[Matching Elasticity ($\frac{d\ln m}{d\ln V}$)]{\includegraphics[width = 0.37\textwidth]
  {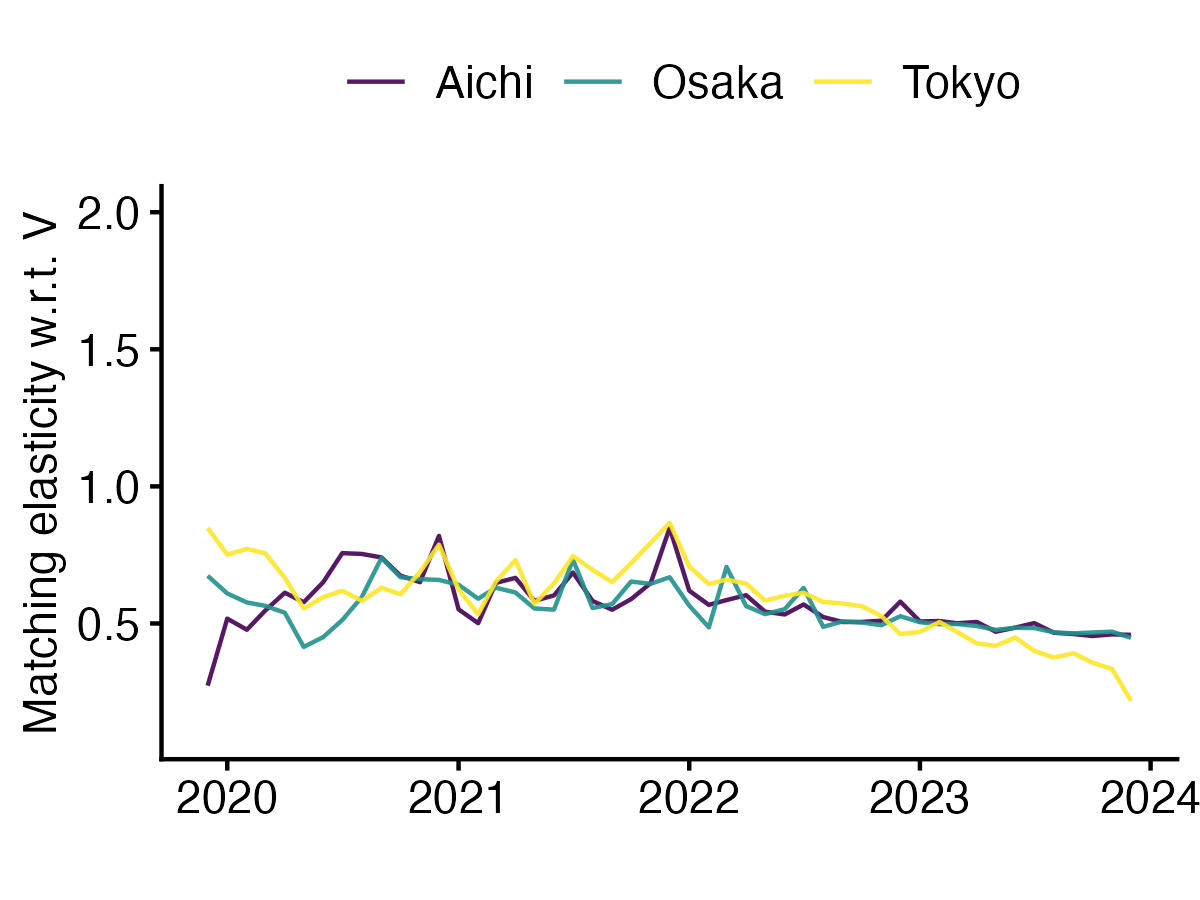}}
  \caption{\textcolor{black}{The heterogeneity across prefectures 2019-2023}}
  \label{fg:matching_efficiency_month_aggregate_each_industry} 
  \end{center}
  \footnotesize
  Note: For confidentiality reasons, we normalize the level of the y-axis in Panels (a)-(c) to the level of Tokyo in December 2019.
\end{figure} 

\textcolor{black}{Figure \ref{fg:matching_efficiency_month_aggregate_each_industry} illustrates prefecture-level labor market trends across the three representative prefectures -- Aichi, Osaka, and Tokyo -- which are the centers of the three metropolitan areas in Japan. This exercise is intended to assess whether the platform exhibits similar matching performance across major regions. We focus on these three prefectures because many smaller prefectures are too thin, especially early in the sample, for spot-work matching to form a meaningful market-level series, and because \cite{kanayama2026justminimumwagehikes} shows that Timee first launched in major metropolitan prefectures such as Tokyo, Osaka, and Aichi.} For confidentiality reasons, the y-axis in the following figures reflects values relative to the baseline set by Tokyo in December 2019. Panels (a) and (b) display user and vacancy levels. Tokyo consistently leads in both categories, with particularly steep growth in vacancy postings after 2022, reflecting rapid labor demand expansion. Panel (c) shows that hires increase in tandem, especially in Tokyo, but Aichi and Osaka also experience steady growth.

\if0
Panel (d) presents trends in matching efficiency. While Tokyo exhibits higher levels of efficiency earlier in the sample, Aichi and Osaka steadily catch up, and by 2024, all three prefectures converge to comparable levels, suggesting a gradual equalization of platform performance across regions. 
\fi
\textcolor{black}{Panel (d) presents trends in matching efficiency. The three prefectures show broadly similar movements from 2020 to 2021, followed by a period of divergence from 2021 to 2023. After 2023, the gap in matching efficiency narrows somewhat, indicating partial re-convergence, although noticeable differences remain near the end of the sample. This suggests that regional heterogeneity in platform matching efficiency became less pronounced later in the sample, but was not fully eliminated.}
\textcolor{black}{Panels (e) and (f) report elasticities with respect to users and vacancies, respectively. Both elasticities remain stable across time and across the three prefectures, generally ranging from 0.3 to 0.8 for user elasticity and from 0.4 to 0.6 for vacancy elasticity. The absence of persistent cross-prefecture gaps indicates limited geographic heterogeneity in matching responsiveness on the platform.}

\section{Conclusion} \label{sec:conclusion}
This study reveals significant differences in labor market dynamics between the private online spot work platform and Hello Work, Japan's public employment service. The private platform demonstrates a rapid expansion in part-time spot employment from December 2019 to December 2023, with a notable increase in registered users and a sharp rise in vacancies, particularly after 2022. We find that the private platform has become increasingly effective in facilitating job matches, achieving a worker finding rate where 80\% of spot jobs are filled, reflecting its growing influence in the part-time labor market.

By employing a novel nonparametric approach, the study estimates the matching function, highlighting clear distinctions in matching efficiency and elasticity between the two platforms. For Hello Work, matching efficiency remains stable around the baseline, while the elasticity with respect to users fluctuates around 0.3--0.5 and the elasticity with respect to vacancies ranges roughly from 0.4 to slightly above 1.0. In contrast, the private platform exhibits higher volatility, with matching efficiency peaking in 2023. The elasticity with respect to the number of users fluctuates between 0.2 and 0.3, while the elasticity concerning vacancies fluctuates between 0.7 and 1.1. These patterns suggest that the private platform's matching process is more vacancy‑driven and systematically more responsive to job demand fluctuations than the public counterpart.

\textcolor{black}{At the prefecture level, the three prefectures exhibit broadly similar movements early in the sample, followed by divergence and partial re-convergence later on. Both user- and vacancy-side elasticities remain stable and close in magnitude across regions, pointing to limited geographic heterogeneity in matching responsiveness on the platform.}

This study offers macroeconomic implications and serves as a foundational step toward understanding labor market dynamics at the micro-level. More broadly, the comparison between Hello Work and the platform provides useful insight for modeling gig spot work within search-and-matching frameworks, particularly by highlighting how repeated matching, app-based search, and differences in measurement shape observed matching outcomes. Future research will explore the behaviors of individuals on both the labor demand and supply sides and their interactions that drive aggregate patterns. This approach will enable a more detailed analysis of the mechanisms underpinning labor market functioning.

\bibliographystyle{ecca}
\bibliography{utmd_timee}

@article{adams2025gender,
  title={The Gender Wage Gap in an Online Labor Market: The Cost of Interruptions},
  author={Adams, Abi and Hara, Kotaro and Milland, Kristy and Callison-Burch, Chris},
  journal={Review of Economics and Statistics},
  volume={107},
  number={1},
  pages={55--64},
  year={2025},
  publisher={MIT Press 255 Main Street, 9th Floor, Cambridge, Massachusetts 02142, USA~…}
}

@article{ando2023discontinuous,
  title={Discontinuous or Kinked Response? Dynamics of Unemployment Insurance and Public Assistance under the {COVID-19} Crisis}, 
  author={Ando, Michihito and Furuichi, Masato and Kanayama, Hayato},
  journal={working paper},
  year={2023}
}

@article{angrist2021uber,
  title={{Uber} versus taxi: A driver’s eye view},
  author={Angrist, Joshua D and Caldwell, Sydnee and Hall, Jonathan V},
  journal={American Economic Journal: Applied Economics},
  volume={13},
  number={3},
  pages={272--308},
  year={2021},
  publisher={American Economic Association 2014 Broadway, Suite 305, Nashville, TN 37203-2425}
}

@book{autor2019studies,
  title={Studies of labor market intermediation},
  author={Autor, David H},
  year={2019},
  publisher={University of Chicago Press}
}

@article{azar2020concentration,
  title={Concentration in {US} labor markets: Evidence from online vacancy data},
  author={Azar, Jos{\'e} and Marinescu, Ioana and Steinbaum, Marshall and Taska, Bledi},
  journal={Labour Economics},
  volume={66},
  pages={101886},
  year={2020},
  publisher={Elsevier}
}

@article{azar2022estimating,
  title={Estimating labor market power},
  author={Azar, Jos{\'e} A and Berry, Steven T and Marinescu, Ioana},
  year={2022},
  journal={National Bureau of Economic Research}
}

@article{banfi2019high,
  title={Do high-wage jobs attract more applicants? Directed search evidence from the online labor market},
  author={Banfi, Stefano and Villena-Roldan, Benjamin},
  journal={Journal of Labor Economics},
  volume={37},
  number={3},
  pages={715--746},
  year={2019},
  publisher={The University of Chicago Press Chicago, IL}
}

@article{bernstein2022matching,
  title={The matching function and nonlinear business cycles},
  author={Bernstein, Joshua and Richter, Alexander W and Throckmorton, Nathaniel A},
  journal={Journal of Money, Credit and Banking},
  year={2022},
  publisher={Wiley Online Library}
}

@article{borowczyk2013accounting,
  title={Accounting for endogeneity in matching function estimation},
  author={Borowczyk-Martins, Daniel and Jolivet, Gr{\'e}gory and Postel-Vinay, Fabien},
  journal={Review of Economic Dynamics},
  volume={16},
  number={3},
  pages={440--451},
  year={2013},
  publisher={Elsevier}
}

@article{brancaccio2020geography,
  title={Geography, transportation, and endogenous trade costs},
  author={Brancaccio, Giulia and Kalouptsidi, Myrto and Papageorgiou, Theodore},
  journal={Econometrica},
  volume={88},
  number={2},
  pages={657--691},
  year={2020},
  publisher={Wiley Online Library}
}

@article{brancaccio2023search,
  title={Search frictions and efficiency in decentralized transport markets},
  author={Brancaccio, Giulia and Kalouptsidi, Myrto and Papageorgiou, Theodore and Rosaia, Nicola},
  journal={The Quarterly Journal of Economics},
  volume={138},
  number={4},
  pages={2451--2503},
  year={2023},
  publisher={Oxford University Press}
}

@article{brown2016boarding,
  title={Boarding a sinking ship? An investigation of job applications to distressed firms},
  author={Brown, Jennifer and Matsa, David A},
  journal={The Journal of Finance},
  volume={71},
  number={2},
  pages={507--550},
  year={2016},
  publisher={Wiley Online Library}
}

@article{buchholz2022spatial,
  title={Spatial equilibrium, search frictions, and dynamic efficiency in the taxi industry},
  author={Buchholz, Nicholas},
  journal={The Review of Economic Studies},
  volume={89},
  number={2},
  pages={556--591},
  year={2022},
  publisher={Oxford University Press}
}

@article{butschek2022motivating,
  title={Motivating gig workers--evidence from a field experiment},
  author={Butschek, Sebastian and Amor, Roberto Gonz{\'a}lez and Kampk{\"o}tter, Patrick and Sliwka, Dirk},
  journal={Labour economics},
  volume={75},
  pages={102105},
  year={2022},
  publisher={Elsevier}
}

@article{castillo2023benefits,
  title={Who benefits from surge pricing?},
  author={Castillo, Juan Camilo},
  journal={Available at SSRN 3245533},
  year={2023}
}

@article{castillo2024matching,
  title={Matching and pricing in ride hailing: Wild goose chases and how to solve them},
  author={Castillo, Juan Camilo and Knoepfle, Dan and Weyl, E Glen},
  journal={Management Science},
  year={2024},
  publisher={INFORMS}
}

@article{chen2019value,
  title={The value of flexible work: Evidence from {Uber} drivers},
  author={Chen, M Keith and Rossi, Peter E and Chevalier, Judith A and Oehlsen, Emily},
  journal={Journal of political economy},
  volume={127},
  number={6},
  pages={2735--2794},
  year={2019},
  publisher={The University of Chicago Press Chicago, IL}
}

@article{cook2021gender,
  title={The gender earnings gap in the gig economy: Evidence from over a million rideshare drivers},
  author={Cook, Cody and Diamond, Rebecca and Hall, Jonathan V and List, John A and Oyer, Paul},
  journal={The Review of Economic Studies},
  volume={88},
  number={5},
  pages={2210--2238},
  year={2021},
  publisher={Oxford University Press}
}

@article{csahin2014mismatch,
  title={Mismatch unemployment},
  author={{\c{S}}ahin, Ay{\c{s}}eg{\"u}l and Song, Joseph and Topa, Giorgio and Violante, Giovanni L},
  journal={American Economic Review},
  volume={104},
  number={11},
  pages={3529--3564},
  year={2014},
  publisher={American Economic Association 2014 Broadway, Suite 305, Nashville, TN 37203}
}

@article{cullen2018gender,
  title={Gender and Sorting in the On-Demand Economy (Working Paper)},
  author={Cullen, Zoe B and Humphries, John Eric and Pakzad-Hurson, Bobak},
    year={2018} 
}

@article{cullen2021outsourcing,
  title={Outsourcing tasks online: Matching supply and demand on peer-to-peer internet platforms},
  author={Cullen, Zo{\"e} and Farronato, Chiara},
  journal={Management Science},
  volume={67},
  number={7},
  pages={3985--4003},
  year={2021},
  publisher={INFORMS}
}

@article{davis2013establishment,
  title={The establishment-level behavior of vacancies and hiring},
  author={Davis, Steven J and Faberman, R Jason and Haltiwanger, John C},
  journal={The Quarterly Journal of Economics},
  volume={128},
  number={2},
  pages={581--622},
  year={2013},
  publisher={MIT Press}
}

@article{faberman2019intensity,
  title={The intensity of job search and search duration},
  author={Faberman, R Jason and Kudlyak, Marianna},
  journal={American Economic Journal: Macroeconomics},
  volume={11},
  number={3},
  pages={327--357},
  year={2019},
  publisher={American Economic Association 2014 Broadway, Suite 305, Nashville, TN 37203-2425}
}

@article{frechette2019frictions,
  title={Frictions in a competitive, regulated market: Evidence from taxis},
  author={Frechette, Guillaume R and Lizzeri, Alessandro and Salz, Tobias},
  journal={American Economic Review},
  volume={109},
  number={8},
  pages={2954--2992},
  year={2019},
  publisher={American Economic Association 2014 Broadway, Suite 305, Nashville, TN 37203}
}

@article{fukai2021describing,
  title={Describing the impacts of {COVID-19} on the labor market in {Japan} until June 2020},
  author={Fukai, Taiyo and Ichimura, Hidehiko and Kawata, Keisuke},
  journal={The Japanese Economic Review},
  volume={72},
  number={3},
  pages={439--470},
  year={2021},
  publisher={Springer}
}

@article{fukui2020job,
  title={Job creation during the {COVID-19} pandemic in {Japan}},
  author={Fukui, Masao and Kikuchi, Shinnnosuke},
  year={2020},
  note = {working paper}
}

@article{gavazza2018aggregate,
  title={Aggregate recruiting intensity},
  author={Gavazza, Alessandro and Mongey, Simon and Violante, Giovanni L},
  journal={American Economic Review},
  volume={108},
  number={8},
  pages={2088--2127},
  year={2018},
  publisher={American Economic Association 2014 Broadway, Suite 305, Nashville, TN 37203}
}

@article{gomme2015worker,
  title={Worker search effort as an amplification mechanism},
  author={Gomme, Paul and Lkhagvasuren, Damba},
  journal={Journal of Monetary Economics},
  volume={75},
  pages={106--122},
  year={2015},
  publisher={Elsevier}
}

@article{guda2019your,
  title={Your{ Uber} is arriving: Managing on-demand workers through surge pricing, forecast communication, and worker incentives},
  author={Guda, Harish and Subramanian, Upender},
  journal={Management Science},
  volume={65},
  number={5},
  pages={1995--2014},
  year={2019},
  publisher={INFORMS}
}

@article{hall2021pricing,
  title={Pricing in designed markets: The case of ride-sharing},
  author={Hall, Jonathan V and Horton, John J and Knoepfle, Daniel T},
  year={2021},
  journal={Working paper, Massachusetts Institute of Technology}
}

@article{hershbein2018recessions,
  title={Do recessions accelerate routine-biased technological change? Evidence from vacancy postings},
  author={Hershbein, Brad and Kahn, Lisa B},
  journal={American Economic Review},
  volume={108},
  number={7},
  pages={1737--1772},
  year={2018},
  publisher={American Economic Association 2014 Broadway, Suite 305, Nashville, TN 37203}
}

@article{higashi2018spatial,
  title={Spatial spillovers in job matching: Evidence from the {Japanese} local labor markets},
  author={Higashi, Yudai},
  journal={Journal of the Japanese and International Economies},
  volume={50},
  pages={1--15},
  year={2018},
  publisher={Elsevier}
}

@article{higashi2025did,
  title={Did {COVID-19} deteriorate mismatch in the {Japanese} labor market?},
  author={Higashi, Yudai and Sasaki, Masaru},
  journal={Japan and the World Economy},
  pages={101331},
  year={2025},
  publisher={Elsevier}
}

@article{kambayashi2006vacancy,
  title={Vacancy market structure and matching efficiency},
  author={Kambayashi, Ry{\=o} and Ueno, Y{\=u}ko},
  year={2006},
  journal={working paper}
}

@article{kambayashi2013role,
  title={The role of public employment services in a developing country: the case of {Japan}in the twentieth century},
  author={Kambayashi, Ryo},
  year={2013},
  journal={PRIMCED Discussion Paper Series 40}
}

@techreport{kambayashi2025decomposing,
  title={Decomposing Recruitment Elasticity in Job Matching},
  author={Kambayashi, Ryo and Kawaguchi, Kohei and Otani, Suguru},
  year={2025},
  institution={IZA Institute of Labor Economics},
  type={IZA Discussion Paper No. 17613},
  url={https://www.iza.org/publications/dp/17613/decomposing-recruitment-elasticity-in-job-matching}
}

@article{kanayama2026justminimumwagehikes,
  title={Just After Minimum Wage Hikes: Short-Run Labor-Demand Response and Reallocation},
  author={Kanayama, Hayato and Miyaji, Sho and Otani, Suguru},
  journal={arXiv preprint arXiv:2505.04555},
  year={2026}
}

@article{kano2005estimating,
  title={Estimating a matching function and regional matching efficiencies: {Japanese} panel data for 1973--1999},
  author={Kano, Shigeki and Ohta, Makoto},
  journal={Japan and the World Economy},
  volume={17},
  number={1},
  pages={25--41},
  year={2005},
  publisher={Elsevier}
}

@article{kassi2018online,
  title={Online labour index: Measuring the online gig economy for policy and research},
  author={K{\"a}ssi, Otto and Lehdonvirta, Vili},
  journal={Technological forecasting and social change},
  volume={137},
  pages={241--248},
  year={2018},
  publisher={Elsevier}
}

@article{katz2019rise,
  title={The rise and nature of alternative work arrangements in the {United States}, 1995--2015},
  author={Katz, Lawrence F and Krueger, Alan B},
  journal={ILR Review},
  volume={72},
  number={2},
  pages={382--416},
  year={2019},
  publisher={SAGE Publications Sage CA: Los Angeles, CA}
}

@article{kawaguchi2021can,
  title={Who can work from home? The roles of job tasks and {HRM} practices},
  author={Kawaguchi, Daiji and Motegi, Hiroyuki},
  journal={Journal of the Japanese and International Economies},
  volume={62},
  number={C},
  year={2021},
  publisher={Elsevier}
}

@article{kawata2021first,
  title={A first aid kit to assess welfare impacts},
  author={Kawata, Keisuke and Sato, Yasuhiro},
  journal={Economics Letters},
  volume={205},
  pages={109928},
  year={2021},
  publisher={Elsevier}
}

@article{kikuchi2021suffers,
  title={Who suffers from the {COVID-19} shocks? Labor market heterogeneity and welfare consequences in {Japan}},
  author={Kikuchi, Shinnosuke and Kitao, Sagiri and Mikoshiba, Minamo},
  journal={Journal of the Japanese and International Economies},
  volume={59},
  pages={101117},
  year={2021},
  publisher={Elsevier}
}

@article{kroft2014does,
  title={Does online search crowd out traditional search and improve matching efficiency? Evidence from {Craigslist}},
  author={Kroft, Kory and Pope, Devin G},
  journal={Journal of Labor Economics},
  volume={32},
  number={2},
  pages={259--303},
  year={2014},
  publisher={University of Chicago Press Chicago, IL}
}

@article{kuhn2004internet,
  title={Internet job search and unemployment durations},
  author={Kuhn, Peter and Skuterud, Mikal},
  journal={American Economic Review},
  volume={94},
  number={1},
  pages={218--232},
  year={2004},
  publisher={American Economic Association}
}

@article{kuhn2013gender,
  title={Gender discrimination in job ads: Evidence from {China}},
  author={Kuhn, Peter and Shen, Kailing},
  journal={The Quarterly Journal of Economics},
  volume={128},
  number={1},
  pages={287--336},
  year={2013},
  publisher={MIT Press}
}

@article{lange2020beyond,
  title={Beyond {Cobb-Douglas}: flexibly estimating matching functions with unobserved matching efficiency},
  author={Lange, Fabian and Papageorgiou, Theodore},
  year={2020},
  journal={National Bureau of Economic Research}
}

@article{lehe2022taxi,
  title={Taxi service with heterogeneous drivers and a competitive medallion market},
  author={Lehe, Lewis and Pandey, Ayush},
  journal={Journal of Urban Economics},
  volume={131},
  pages={103488},
  year={2022},
  publisher={Elsevier}
}

@article{marinescu2018mismatch,
  title={Mismatch unemployment and the geography of job search},
  author={Marinescu, Ioana and Rathelot, Roland},
  journal={American Economic Journal: Macroeconomics},
  volume={10},
  number={3},
  pages={42--70},
  year={2018},
  publisher={American Economic Association 2014 Broadway, Suite 305, Nashville, TN 37203-2425}
}

@article{marinescu2020opening,
  title={Opening the black box of the matching function: The power of words},
  author={Marinescu, Ioana and Wolthoff, Ronald},
  journal={Journal of Labor Economics},
  volume={38},
  number={2},
  pages={535--568},
  year={2020},
  publisher={The University of Chicago Press Chicago, IL}
}

@article{mas2020alternative,
  title={Alternative work arrangements},
  author={Mas, Alexandre and Pallais, Amanda},
  journal={Annual Review of Economics},
  volume={12},
  number={1},
  pages={631--658},
  year={2020},
  publisher={Annual Reviews}
}

@article{matzkin2003nonparametric,
  title={Nonparametric estimation of nonadditive random functions},
  author={Matzkin, Rosa L},
  journal={Econometrica},
  volume={71},
  number={5},
  pages={1339--1375},
  year={2003},
  publisher={Wiley Online Library}
}

@article{miyamoto2025macroeconomic,
  title={Macroeconomic facts in the {Japanese} labor market: survey},
  author={Miyamoto, Hiroaki},
  journal={The Japanese Economic Review},
  pages={1--46},
  year={2025},
  publisher={Springer}
}

@article{mukoyama2018job,
  title={Job search behavior over the business cycle},
  author={Mukoyama, Toshihiko and Patterson, Christina and {\c{S}}ahin, Ay{\c{s}}eg{\"u}l},
  journal={American Economic Journal: Macroeconomics},
  volume={10},
  number={1},
  pages={190--215},
  year={2018},
  publisher={American Economic Association 2014 Broadway, Suite 305, Nashville, TN 37203-2425}
}

@article{otani2024nonparametric,
  title={Nonparametric Estimation of Matching Efficiency and Mismatch in Labor Markets via Public Employment Security Offices in {Japan}, 1972-2024},
  author={Otani, Suguru},
  journal={arXiv preprint arXiv:2407.20931},
  year={2024}
}

@article{otani2025marriage,
title = {Nonparametric estimation of matching efficiency and elasticity in a marriage agency platform: 2014–2025},
journal = {Economics Letters},
volume = {256},
pages = {112617},
year = {2025},
issn = {0165-1765},
doi = {https://doi.org/10.1016/j.econlet.2025.112617},
url = {https://www.sciencedirect.com/science/article/pii/S0165176525004549},
author = {Suguru Otani}
}

@article{otani2025onthejob,
  title={Nonparametric estimation of matching efficiency and elasticity on a private on-the-job search platform: Evidence from {Japan}, 2014-2024},
  author={Otani, Suguru},
  journal={Journal of the Japanese and International Economies},
  pages={101394},
  year={2025},
  publisher={Elsevier}
}

@article{petrongolo2001looking,
  title={Looking into the black box: A survey of the matching function},
  author={Petrongolo, Barbara and Pissarides, Christopher A},
  journal={Journal of Economic literature},
  volume={39},
  number={2},
  pages={390--431},
  year={2001},
  publisher={American Economic Association}
}

@book{pissarides2000equilibrium,
  title={Equilibrium unemployment theory},
  author={Pissarides, Christopher A},
  year={2000},
  publisher={MIT press}
}

@article{rogerson2005search,
  title={Search-theoretic models of the labor market: A survey},
  author={Rogerson, Richard and Shimer, Robert and Wright, Randall},
  journal={Journal of economic literature},
  volume={43},
  number={4},
  pages={959--988},
  year={2005},
  publisher={American Economic Association}
}

@article{rosaia2020competing,
  title={Competing platforms and transport equilibrium: Evidence from {New} {York} {City}},
  author={Rosaia, Nicola},
  journal={Unpublished Manuscript URL https://economics. sas. upenn. edu/system/files/2021-09/Rosaia. pdf},
  year={2020}
}

@article{sasaki2008matching,
  title={Matching Function For The {Japanese} Labour Market: Random Or Stock--Flow?},
  author={Sasaki, Masaru},
  journal={Bulletin of Economic Research},
  volume={60},
  number={2},
  pages={209--230},
  year={2008},
  publisher={Wiley Online Library}
}

@article{sedlavcek2014match,
  title={Match efficiency and firms' hiring standards},
  author={Sedl{\'a}{\v{c}}ek, Petr},
  journal={Journal of Monetary Economics},
  volume={62},
  pages={123--133},
  year={2014},
  publisher={Elsevier}
}

@article{sekiya2026concentration,
  title={Concentration Control in Recommendations: Algorithm Design and Field Experiment on a Spot Gig-Work Platform},
  author={Sekiya, Kazuki and Otani, Suguru and Komatsu, Yuki and Fujii, Yuki and Ozeki, Shunsuke and Noda, Shunya},
  journal={Working Paper},
  year={2026}
}

@article{shibata2020labor,
  title={Is Labor Market Mismatch a Big Deal in {Japan}?},
  author={Shibata, Ippei},
  journal={The BE Journal of Macroeconomics},
  volume={20},
  number={2},
  pages={20160179},
  year={2020},
  publisher={De Gruyter}
}

@article{wolthoff2014s,
  title={It's about time: implications of the period length in an equilibrium search model},
  author={Wolthoff, Ronald},
  journal={International Economic Review},
  volume={55},
  number={3},
  pages={839--867},
  year={2014},
  publisher={Wiley Online Library}
}

\end{document}